\documentclass{aa}  

\usepackage{graphicx}
\usepackage{txfonts}
\usepackage{enumitem}
\usepackage{amsmath,amssymb}
\usepackage{mathtools}
\usepackage{textcomp}
\usepackage{natbib}
\usepackage{tikz}  
\usetikzlibrary{arrows,snakes,shapes,fadings}
\usetikzlibrary{arrows.meta}
\usepackage{ulem}  
\usepackage{float}
\usepackage{afterpage}
\usepackage{gensymb}

\newcommand{\ond}{Ond\v{r}ejov}
\newcommand{\kms}{km~s$^{-1}$}

\newcommand{\mlr}{$\dot{M}$}

\begin{document}

   \title{First detection of a magnetic field in low-luminosity B[e] stars
      \thanks{Based on data from Perek 2~m \, telescope, Ond\v{r}ejov, Czech Republic. \newline
       \hspace*{3mm} Based on observations obtained at the Canada-France-Hawaii Telescope
      (CFHT) which is operated by the National Research Council of Canada,
      the Institut National des Sciences de l'Univers of the Centre National
      de la Recherche Scientique of France, and the University of Hawaii.
      The observations at the Canada-France-Hawaii Telescope were performed
      with care and respect from the summit of Maunakea which is a significant
      cultural and historic site.\newline
      }
      \thanks{This work uses NIST database \cite*{NIST_ASD}. NIST Atomic Spectra Database (ver. 5.6.1), [Online]. 
        Available: https://physics.nist.gov/asd [2019, May 25]. National Institute of Standards and Technology, Gaithersburg, MD. DOI: https://doi.org/10.18434/T4W30F, and  
       \cite{Peters_line_list} line list and database.}
     }

   \subtitle{New scenarios for the nature and evolutionary stages of  FS CMa stars}

   \author{
    D. Kor\v{c}\'akov\'a\inst{\ref{inst_P}}
     \and
    F. Sestito\inst{\ref{inst_Victoria}}
     \and
    N.~Manset\inst{\ref{inst_CFHT}}
    \and
    P.~Kroupa\inst{\ref{inst_P},\ref{inst_Bonn}}
     \and
    V.~Votruba\inst{\ref{inst_B}}
      \and
    M.~\v{S}lechta\inst{\ref{inst_O}} 
      \and
    S.~Danford \inst{\ref{inst_G}}
       \and
    N.~Dvo\v{r}\'{a}kov\'{a}\inst{\ref{inst_P}}
    \and
    A.~Raj\inst{\ref{inst_Delphi},\ref{inst_I}}
       \and
       S.\,D.~Chojnowski\inst{\ref{inst_montana}}
       \and
       H.\,P.~Singh \inst{\ref{inst_Delphi}}
          }
   \institute{
    Charles University, Faculty of Mathematics and Physics, Astronomical Institute, V Hole\v{s}ovi\v{c}k\'ach 2, CZ-180 00 Praha 8, Czech Republic
    \email{kor@sirrah.troja.mff.cuni.cz}\label{inst_P} 
     \and Department of Physics and Astronomy, University of Victoria, Victoria, BC, V8W 3P2, Canada \label{inst_Victoria}
     \and Canada-France-Hawaii Telescope Corporation, 65-1238 Mamalahoa Hwy, Kamuela HI 96743\label{inst_CFHT} 
     \and Helmholtz-Institut f\"{u}r Strahlen- und Kernphysik, University of Bonn, Nussallee 14-16, D-53115 Bonn, Germany\label{inst_Bonn} 
     \and Institute of Theoretical Physics and Astrophysics, Masaryk 
          University, CZ-611 37 Brno, Kotl\'a\v{r}sk\'a 2, Czech Republic\label{inst_B} 
     \and Astronomical Institute of the~Academy of Science of the~Czech 
          Republic, Fri\v{c}ova 298, CZ-251 65 Ond\v{r}ejov, Czech Republic\label{inst_O} 
     \and Department of Physics and Astronomy, University of North Carolina at Greensboro, Greensboro, NC 27402, USA \label{inst_G}
     \and Department of Physics and Astrophysics, University of Delhi, Delhi 110007, India\label{inst_Delphi}
     \and Indian Institute of Astrophysics, Block II, Koramangala, Bangalore 560034, India\label{inst_I}
     \and Department of Physics, Montana State University, P.O. Box 173840, Bozeman, MT 59717--3840, USA\label{inst_montana}
   }

   \date{Received May 8, 2021; accepted December 31, 2021}

  \abstract{
We report the first detection of the magnetic field in a star of FS~CMa type, a subgroup of objects characterized by the B[e] phenomenon. 
The split of magnetically sensitive lines in IRAS 17449+2320 determines the magnetic field modulus of $6.2\pm 0.2$~kG. 
Spectral lines and their variability reveal the presence of a B-type spectrum and a hot continuum source in the visible.
The hot source confirms GALEX UV photometry. Because there is a lack of spectral lines for the hot source
in the visible, the spectral fitting gives only the lower temperature limit of the hot source, which is 
50~000~K, and the upper limit for the B-type star of 11~100~K. The $V/R$ ratio of the H$\alpha$ line shows quasiperiodic 
behavior on  timescale of 800 days. We detected a~strong red-shifted absorption in the wings of Balmer 
and \ion{O}{i} lines in some of the spectra. The absorption lines of helium and other metals show no, or very small, variations, 
indicating unusually stable photospheric regions for FS~CMa stars. We detected two events of 
material infall, which were revealed to be discrete absorption components of resonance lines.

The discovery of the strong magnetic field together with the Gaia measurements of the proper motion show
that the most probable nature of this star is that of a post-merger object created after the leaving the binary 
of the birth cluster. Another possible scenario is a magnetic Ap star around Terminal-Age Main Sequence (TAMS). On the other hand, the strong magnetic 
field defies the hypothesis that IRAS 17449+2320 is an extreme classical Be star.  Thus, IRAS 17449+2320 provides a pretext for exploring 
a new explanation of the nature of FS~CMa stars or, at least, a~group of stars with very similar spectral properties. 
}

   \maketitle
%

\section{Introduction}

IRAS 17449+2320 (BD +23 3183) belongs to a peculiar B-type stellar group known as FS CMa stars, which includes only about 
sixty members, together with the candidate objects. These stars show the B[e] phenomenon first described by \cite{Allen76}.
Further detailed research and a classification of the B[e] stars was carried out by \cite{Lamers98}, who recognized that the 
B[e] phenomenon, namely, the presence of forbidden lines and infrared excess, includes stars of different types and 
at different evolutionary stages. They found the B[e] stars among supergiants, compact planetary nebulae,
Herbig Ae/Be stars, and symbiotic stars. However, they were not able to classify about half of stars known at that time. 
Later, \cite{Miroshnichenko_FS_CMa} noted that almost all unclassified stars share a set of properties in common and introduced 
a~new group called FS~CMa stars. He found the following common signatures:  Balmer lines have stronger emission than 
what is observed in classical Be stars; emission lines of neutral or singly ionized metals from permitted as well as 
forbidden transitions are also present; weak emission lines of [\ion{O}{iii}] may be detected; the infrared excess 
has a peak around 20 $\mu$m and sharply decreases toward the longer wavelengths; FS~CMa stars are located outside 
star-forming regions; their temperatures are in the range between 9\,000 and 30\,000~K and luminosity $\log L/L_{\odot}$ 2.5 and 4.5. 
The FS~CMa group was extended and described in \cite{Miroshnichenko07_II, Miroshnichenko11, Miroshnichenko_2017_FS_CMa}.

Forbidden lines and infrared excess are indicators of extended circumstellar matter, which complicates
the study of the central object. In some cases, we have no information drawn directly from the photosphere. Despite this obstacle, 
it has nonetheless been found that FS~CMa stars are near the Terminal-Age Main Sequence (TAMS) in the Hertzsprung-Russell diagram \citep{Miroshnichenko_FS_CMa,Miroshnichenko17, Miroshnichenko20_3Pup}. 
The presence of circumstellar matter significantly changes the spectrum of the central object. In particular, the UV region may be affected,
depending on the angle of view, by the strong absorption of iron group elements, referred to as the so-called ``iron curtain''. 
A great number of absorption lines creates a ``false continuum'' and may reduce the outgoing UV flux by about an order
of magnitude \citep{Ivan-Pariz}. The energy absorbed in the UV lines is reradiated in the visible and near IR due 
to the cascade process. Such lines exhibit broad emission wings. The visible spectrum is dominated
by very strong emission in the H$\alpha$ line. The H$\alpha$ line may be as much as a hundred times brighter than the continuum in some cases. 
The strongest forbidden lines are [\ion{O}{i}] $\lambda \lambda$ 6~300, and 6~364~\AA, which are always present. Usually 
 [\ion{S}{ii}] $\lambda \lambda$ 6~716, 6~731~\AA\, doublet is detected and in some objects also the doublet of 
[\ion{N}{ii}] $\lambda \lambda$ 6~548, 6~583~\AA. Resonance lines, \ion{Na}{i} D1, D2, and \ion{Ca}{ii} H, and K lines, show a~broad
emission. The He lines and permitted metal lines may be seen in the emission as well as in the absorption. They usually show rapid night-to-night
variability with extreme changes of the line profile from the pure absorption to the P-Cygni profile, inverse P-Cygni profile, 
pure emission, or absorption with the emission wings (HD~50138, \citealt{Pogodin97}; MWC~342, \citealt{Blanka}).

Currently, there is sufficient spectroscopic and photometric data 
for some FS~CMa stars to search for variability from hours to decades
(MWC~623, \citealt{Honza12};
MWC~342, \citealt{Blanka};
MWC~728, \citealt{MWC728};
HD~50138, \citealt{Terka17};
HD~85567, \citealt{Khokhlov17};
AS~386, \citealt{Khokhlov18};
3~Pup, \citealt{Miroshnichenko20_3Pup}).
Other studies are based on short observation runs. It was found that spectral lines of most stars show variability 
at different timescales. Permitted absorption lines of He and metals
show usually night-to-night variability, while forbidden emission lines show changes on timescales of months 
or years. The periodicity of the H$\alpha$ line has been found on the order of several weeks to years. The best studied 
photometric periodicity is for MWC~342 \citep{Shevchenko93,MelNikov97,Chkhikvadze02}. It shows a short period between 
14 and 16 days, with a longer one around 40 and 120 days. The latter has not been found in every season, even if 
the data would be sufficient to show it. The change of the period from season to season is likely to be a real effect. 
 Here, we note that the scale of the variability ought to be considered, rather than the regular periodicity. On the other hand, 
a~short period of around 27.5~d connected with the orbital motion of the binary has been found for MWC~728 \citep{MWC728}.

The variability of the line profiles reveals that several physical phenomena play an important role in the envelopes of FS~CMa stars.
The constant  outflow of material is accompanied by the appearance of expanding layers \citep[in MWC~342,][]{Blanka} 
that may even slow down. The variable shell structure was proven interferometrically by \cite{Kluska16}.
Moreover, the episodic material ejecta or cause infall are observable in the resonance lines as discrete absorption
components \citep{FSCMa_diag}. Inhomogenities in a~rotating disk were discovered in HD~50138 \citep{Terka17}. 

There have been several attempts to measure the mass-loss rate ($\dot{M}$) in FS~CMa stars:
HD~87643 \citep{Pacheco82}, AS~78 \citep{Miroshnichenko_AS_78}, and IRAS~00470+6429 \citep{Carciofi10}. The 
measured values are in the range of $2.5 \cdot 10^{-7}$ to $1.5 \cdot 10^{-6} M_{\odot} \cdot yr^{-1}$. 
In particular, \citet{Pacheco82} noted that such a large mass loss cannot be reached by the radiatively driven wind 
of the central star. This finding is taken as the strongest argument in support of the binarity of FS~CMa stars. The 
large amount of circumstellar matter is naturally explained in this case as resulting from prior or current mass transfer events. 
A~detailed discussion of binarity is presented by 
\cite{Miroshnichenko15_dvojhvezdy, 2028_Cyg_bisector, Miroshnichenko_20_binarity}. Indeed, the periodicity, which is interpreted 
as the orbital motion, was found in a~significant stellar sample. To support the binary hypothesis, there are a few examples among FS~CMa
stars which display a composite spectrum: these are the hot B-type as well as late, usually K-type components. This may be taken as a definitive proof of 
binarity. However, we are dealing with stars surrounded by a huge amount of circumstellar matter forming the geometrically thick disk and, thus, the interpretation is not so straightforward in this case. The K-type spectrum may be formed through the disk, which behaves 
as a pseudoatmosphere. The analysis and modeling of the H$\alpha$ bisector variability of MWC~623 \citep{2028_Cyg_bisector} 
favors the radiative transfer effect through the circumstellar disk. The result of this modeling is supported 
by the polarimetric observations of \cite{Zickgraf89}, which indicates an edge-on view. Up to now, no straight spectral disentangling
of any FS~CMa star has been successful. The binarity of most (if not every) FS~CMa star is a realistic 
hypothesis because it has been found that most hot stars are indeed part of binaries or multiple systems. A~more 
difficult question to tackle is how the binarity is connected with the observed properties of FS~CMa stars. 

At this point, we reach back to the three mass-loss rate calculations of \cite{Pacheco82}, \cite{Miroshnichenko_AS_78}, and \cite{Carciofi10}. 
There is a serious hidden problem, as every technique is slightly different, but the common factor is the assumption of 
a smoothly accelerated wind, according to the $\beta$ velocity law or similar. When these calculations were performed, 
there were no extended observation campaigns and the authors had no information about the velocity field structure in the envelopes. 
We now know that the expanding layers may be decelerated \citep{Blanka}. The accumulation of the matter around a~star leads to 
an overestimation of the \mlr \, using the smoothly accelerated velocity law. Therefore the real \mlr \, may be at least 
an order of magnitude smaller. Moreover, all three stars are massive FS~CMa stars, namely, the radiation pressure is larger 
than in other members of the group. The inappropriate usage of the velocity law leads us to discard the strongest argument for 
the binarity of FS~CMa stars. Evidence against the binary hypothesis also comes from the interferometric observations of \cite{Kluska16}. 
The observations are better described by the model of a disk with a spot rather than by a~companion star.

All observed phenomena fit merger, or post-merger, objects. We discuss the  pro and con arguments in the Sect.~\ref{mergers}. 
The first observation, which points to this hypothesis published \cite{Fuente15}. They found two FS~CMa stars in the central 
parts of two clusters. 

Our first spectrum of IRAS 17449+2320 indicated that this star may play an important role in revealing the nature of 
the FS~CMa stars. We started the observation campaign at the Ond\v{r}ejov observatory, which was later extended 
at other observatories. The results of these data and older archival data are presented in this paper. 

\section{Observations, data reduction, and the line identification}
\label{Observations}

\subsubsection*{Observations and data reduction}

We were able to collect data from 2005 to 2019 from eight observatories: 
Canada France Hawaii Telescope, Hawai, USA (CFHT);
Apache Point Observatory, New Mexico, USA (APO);
Himalayan Chandra Telescope, Leh-Ladakh, India (HCT);
Observatorio Astronomico Nacional San Pedro Martir, Baja California, Mexico (SPM);
Tien-Shan Astronomical Observatory, Almaty, Kazakhstan (TSAO); 
Ond\v{r}ejov Observatory, Czech Republic (OO);
Three College Observatory, North Carolina, USA (TCO), and
Bellavista Obs. L, Italy (BO).
The diameter of the primary mirrors of these telescopes and resolving 
powers of the individual spectrographs are summarized in Table~\ref{param}.

The~data\footnote{Spectra are available at the CDS database.} are reduced in 
\texttt{IRAF}\footnote{ 
  \texttt{IRAF} is distributed by the~National Optical Astronomy Observatories, 
  operated by the~Association of Universities for Research in Astronomy, 
  Inc., under contract to the~National Science Foundation of the~United States.
}
using IRAF's standard procedures, only SPMO data are without the flat field correction.
The SPMO and OO spectra are readout without the optimal extraction and the program \texttt{dcr} \citep{Pych}
is applied to OO data to remove cosmic rays. The detailed information of the spectra are in 
Table \ref{sptab}. 

\begin{table}[!t]
  \caption{Parameters of the spectrographs used.}
  \label{param}
  \centering
  \begin{tabular}{lrll}
  \hline \hline
                     & R (H$\alpha$) $\approx$  & spectrograph & PMD  \\
                     &                          &         & (m)  \\ 
  \hline
CFHT                 & 65\,000   &  ESPaDOnS   & 3.6 \\
APO                  & 31\,500   &  ARCES      & 3.5 \\ 
HCT                  & 30\,000   & HESP        & 2.01 \\
SMPO                 & 18\,000   & \`{e}chelle & 2.12  \\
TSAO                 & 15\,500   & \`{e}chelle & 1.0 \\ 
OO                   & 12\,500   & slit        & 2.0 \\
TCO                  & 12\,000   & \`{e}chelle & 0.8 \\
BO                   &  \,\,600 \& 5\,200  &  LHIRES3  &  0.235\\ 
 \hline
 \end{tabular}
  \tablefoot{
    The resolution, type of the spectrograph, and the diameter of the primary mirror are summarized. \\
   }
\end{table}

\subsubsection*{Line identifications}
\label{line_identification_sec}

The FS CMa stars have complicated spectra. The spectrum may exhibit lines from a hot and a cool component as well as 
those formed in the circumstellar medium. Accurate line identification and position determinations are crucial
for analysis and modeling. Therefore, we decided to do the line identification manually following the lines from multiplets. 
We present the list of identified lines in the appendix \ref{linestab}. We also plotted parts of the spectrum with 
\ion{He}{i} lines (Fig.~\ref{He_identification}), which are important for determination of the spectral type 
because the spectrum shows signatures of a late B- or early A-type star. Besides the strong Balmer and Pashen 
hydrogen lines and weak \ion{He}{i} lines, we found lines of 
\ion{C}{i},
\ion{N}{i},
\ion{O}{i},
\ion{Na}{i},
\ion{Mg}{i},
\ion{Mg}{ii},
\ion{Si}{i},
\ion{Si}{ii},
\ion{Ca}{i},
\ion{Ca}{ii},
\ion{Ti}{ii},
\ion{Cr}{ii},
\ion{Mn}{i},
\ion{Fe}{i}, and
\ion{Fe}{ii}.
All the metal lines are in the absorption, with the exception of the [\ion{O}{i}] $\lambda \lambda$ 
5~577, 6~300, and 6~364~\AA\, lines and very weak autoionization \ion{O}{i} lines at $4355 -4356$~\AA.

\section{Spectral properties}

The most distinctive feature in the spectrum of IRAS~17449+2320 is the H$\alpha$ line.
Its intensity, reaching values from five to ten times the continuum level, along with the line profile,
showing the absorption wings, as well as the $V/R$ variability indicate that we are dealing with a very special object. 
The forbidden lines are relatively weak, and we have been able to detect only
the forbidden lines of neutral oxygen 5~577~\AA, 6~300, and 6~364~\AA. 

The star is very bright in the UV region compared to a classical Be star (see Sect.~\ref{UV_sec}). The UV radiation affects 
the properties of the spectral lines in the circumstellar region. The lines, which are radiatively connected with 
the resonance lines, show broad emission wings over the absorption core. However, this non-LTE effect is probably 
not as strong in IRAS~17449+2320 as it is in other FS~CMa stars \citep{FSCMa_diag}. This kind of a line profile 
is detected only in a few spectral lines. Therefore, we can expect no (or a very weak) iron curtain in the UV region 
of its spectrum. The high energy of photons also allows for the creation of autoionization lines, which we were able 
to detect in neutral oxygen as very weak emission and absorption lines. However, we found no Raman lines. 

The spectrum is contaminated by interstellar 
absorption in the resonance lines \ion{Na}{i} D1, D2, \ion{Ca}{ii} H, and K, 
as well as diffuse interstellar bands (DIBs). The strongest DIB is at 6~614~\AA. 
Two other bands,  5~780, and 5~797~\AA, have almost the same intensity.
Following the DIB's families \citep{Galazutdinov00, DIB_families}, 
we also identified DIBs at 6~196 and 6~379~\AA.

\subsection{UV radiation}
\label{UV_sec}

The information about the UV region is limited to data from the GALEX mission. Table~\ref{GALEX_tab} summarizes the only 
observation for IRAS~17449+2320. We used the classical Cardelli's law \citep{Cardelli} for the dereddening 
with the parameter $R_{v}=3.1$. The GALEX catalog gives E(B-V)= 0.083, which is based on the map of  \cite{Schlegel98}.
We take the limiting values of E(B-V) as 0.04 and 0.083 corresponding to the spectral types A0 and 
B9 (see Sect.~\ref{spectral_type} for details), and the intrinsic brightness is in the range $12.2 - 11.9$ mag for FUV, and 
 $12.1 - 11.7$~mag for NUV GALEX bands. 

Based on the Gaia parallax (Table~\ref{gaia_data}) the absolute brightness in NUV is between 2.7 and 2.4~mag.
This value can be compared with that of a classical Be star, which was observed by GALEX. One of such stars with similar 
parameters ($T_{\rm{eff; Gaia}}$, B9IVn) is 2~Cet. According to the Gaia DR3 parallax and GALEX E(B-V) index, its absolute 
brightness in the NUV filter is 4.45~mag. If we take the maximum  E(B-V)=0.083 to obtain the upper limit for the NUV brightness 
of 2 Cep, we obtain the value of 3.9~mag. We note that Be stars were not usually observed in the FUV band. The uncertainties 
in observation can not explain such a~huge difference and IRAS~17449+2320 does have an UV excess compared to the single star without 
a~magnetic field. As we show in the next sections, this UV excess plays an important role in spectrum formation as well as 
the determination of the stellar parameters.
\begin{table}[h!]
\caption[]{GALEX data} 
\label{GALEX_tab}
\begin{tabular}{lccc}
\hline \hline
                    & brightness          & intrinsic  &  flux      \\
                    &                     & brightness & \\
                    &   (mag)             & (mag)                &    ($\mu$J) \\ \hline
FUV  &  $12.527 \pm 0.004$ & 11.847 & $35400  \pm 112$   \\ 
NUV  &  $12.485 \pm 0.002$ & 11.720 & $36800  \pm 67 $    \\ \hline 
\end{tabular}
\tablefoot{
The intrinsic brightness is calculated for the value of E(B-V) tabulated in the GALEX catalog ($\sim$0.083) and
errors are commented in the text.
The FUV camera wavelength region is from 1\,344 to 1\,786 \AA, and NUV region from 1\,771 to  2\,831~\AA.
}
\end{table}

\subsection{Hydrogen lines}

The Balmer series lines that were detected are very strong (see Fig.~\ref{variability_component}, or 
\ref{variability_night-to-night}). We found 17 Balmer lines (Apache Point Observatory, 2016/06/16).
The higher members of the series are found fully in the absorption. The first line where a weak
emission component appears is H$\theta$. The emission is shifted bluewards by $\sim$80 \kms \, from the center of the 
line. The lower the member of the series, the emission is stronger and broader. This behavior, connected with the line formation, 
is seen in other hot emission-line stars. Beginning about H$\delta$ the central emission becomes more complicated
(see Fig.~\ref{variability_component}, or \ref{variability_night-to-night}). The emission in the H$\alpha$ line reaches about 
5 -- 10~x the continuum level. The double-peaked asymmetric emission overlaps the broad absorption wings. The line shows strong
$V/R$ variations (Sect.~\ref{variability}).

Our optical spectra contain the Paschen series lines starting from the Pa $\epsilon$. The latter shows a broad absorption component
with a very narrow and strong central emission. The higher the member of the series, the stronger the emission relative 
to the absorption (see Fig.~\ref{paschen}). The lines from the 18th level are seen fully in emission. We were able 
to find the line from the 28th level. This sets an upper limit of the electron density at $\sim 2\cdot10^{12}$ and 
of the turbulent velocity at $\sim$ 130~\kms \citep{Inglis_Teller, Nissen}.

\subsection{Helium lines}

He lines are represented by relatively strong triplet states (4~026, 4~471, 4~713, and 5~876~\AA)
and weak singlet states (4~922, 5~016, and 6~678~\AA), as shown in Fig.~\ref{He_identification}.
The line 6~678~\AA \, shows strong emission wings that are also detectable in  5~876~\AA \,  line.
These two are the only  \ion{He}{i} lines showing a variability. The other \ion{He}{i} lines are seen purely in absorption and
are remarkably stable (Fig.~\ref{He_variability_fig}). This is the opposite behavior to what is generally observed in other FS~CMa stars. 
Typically, the absorption lines show rapid night-to-night changes, especially \ion{He}{i} lines. The line shape may show 
changes from pure absorption to the P~Cygni profile, inverse P~Cygni profile, or emission wings or even pure emission.

\subsection{Oxygen lines}

We detected only the spectral lines of neutral oxygen in the spectrum of IRAS~17449+2320. 
The strongest are the triplet at 7~772, 7~774, and 7~775~\AA\, and the line at 8~446~\AA.
The triplet shows a wide emission overlapped by very strong and relatively sharp absorption of the individual
components. All three lines are Zeeman-split. The 8~446~\AA\, line is also in emission with a very narrow and 
deep absorption component, which in some spectra, reaches intensities that are below the continuum level. 

There is also a strong [\ion{O}{i}] doublet at $\lambda\lambda\, 6~300, 6~364$~\AA. Both lines are purely in emission and 
symmetric. Our best spectrum, which has the highest resolution and highest $S/N$ (CFHT 2017-08-14), shows a hint of a very low split blue-shifted emission 
in the $6~364$~\AA\, line. Unfortunately, it is impossible to confirm this suggestion on the $6~300$~\AA \, line because of a contamination with telluric lines. 
The position of the [\ion{O}{i}] doublet does not vary. We found an average value of $-16 \pm 2$~\kms, which is consistent
with the measurement of \cite{Aret16}. This allows us to use the radial velocities (RVs) of the [\ion{O}{i}] as a reference 
system value. There is another forbidden line from the same deexcitation cycle (Fig.~\ref{grotrian_Federico}) at 5~577~\AA.

Other lines are weak and observed in absorption (Table~\ref{linestab}) with the exception of a weak broad emission formed by the
semiforbidden lines in $4~355 - 4~356$~\AA \, with upper levels above the ionization threshold. Other oxygen autoionization
lines present in the spectrum have been found at 3~952, 4~918, 5~573, and 7~157~\AA.

\subsection{Resonance lines}
\label{resonance_sec}

\begin{figure}[h!]
  \resizebox{\hsize}{!}{\includegraphics{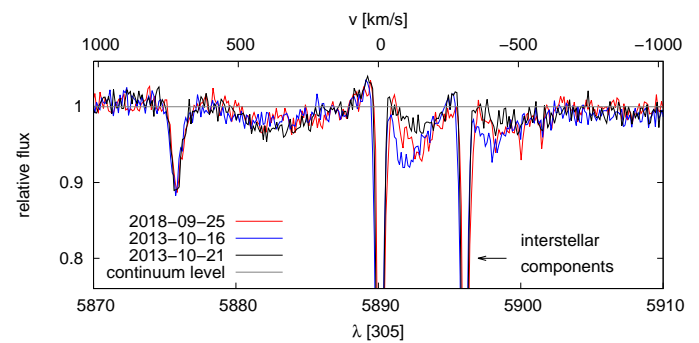}}
  \caption{
    Discrete components of the resonance lines. The events were captured on the SPM spectra on 2013-10-16, following the spectrum 
    over five days, and on 2018-09-25.
   }
  \label{resonance_discrete}
\end{figure}

Only two resonance doublets, \ion{Na}{i}~D1,~D2, and \ion{Ca}{ii} H, and K lines, have been detected in the wavelength range from about 3~700 to 10~000~\AA.
The lines always show a broad emission ($\pm 100$~\kms) overlapped by interstellar components. In our spectra, we were 
able to detect two events of the material ejecta or infall revealing itself as discrete absorption components of these resonance 
lines (2013-10-16 and 2018-09-25, Fig.~\ref{resonance_discrete}). The red absorption at 2013-10-16 is shifted 
about 85~\kms relative to the system velocity determined by [\ion{O}{i}] lines. 
Five days later, the absorption was significantly shallower and at the position around 130~\kms also redward. 
This acceleration of the material implies the material infall rather than ejection.

The resonance \ion{Li}{i} doublet at 6~708~\AA\, deserves a special note, although we were unable to detect it in any of our spectra.
\ion{Li}{i} lines have been detected in about half of the FS~CMa stars \citep{lithium_conference}. 
Their presence has been considered a~proof of binarity \citep[e.g.,][]{MWC728} because of the low ionization potential of \ion{Li}{i}.
However, some signatures of these lines can be interpreted as having been formed in the circumstellar disk. 
The lack of \ion{Li}{i} resonance lines  may therefore point to the near pole-on orientation.
However, this hypothesis has to be taken with caution. The lack of \ion{Li}{i} resonance lines may
be explained in different ways, especially the optical depth along the line-of-sight is crucial here.
Nevertheless, the possibility that we see the system almost pole-on should be taken into account.

\section{Variability}
\label{variability}

\subsection{Appearence of a red absorption}
\label{red_absorption}

\begin{figure*}[h!]
  \resizebox{\hsize}{!}{\includegraphics{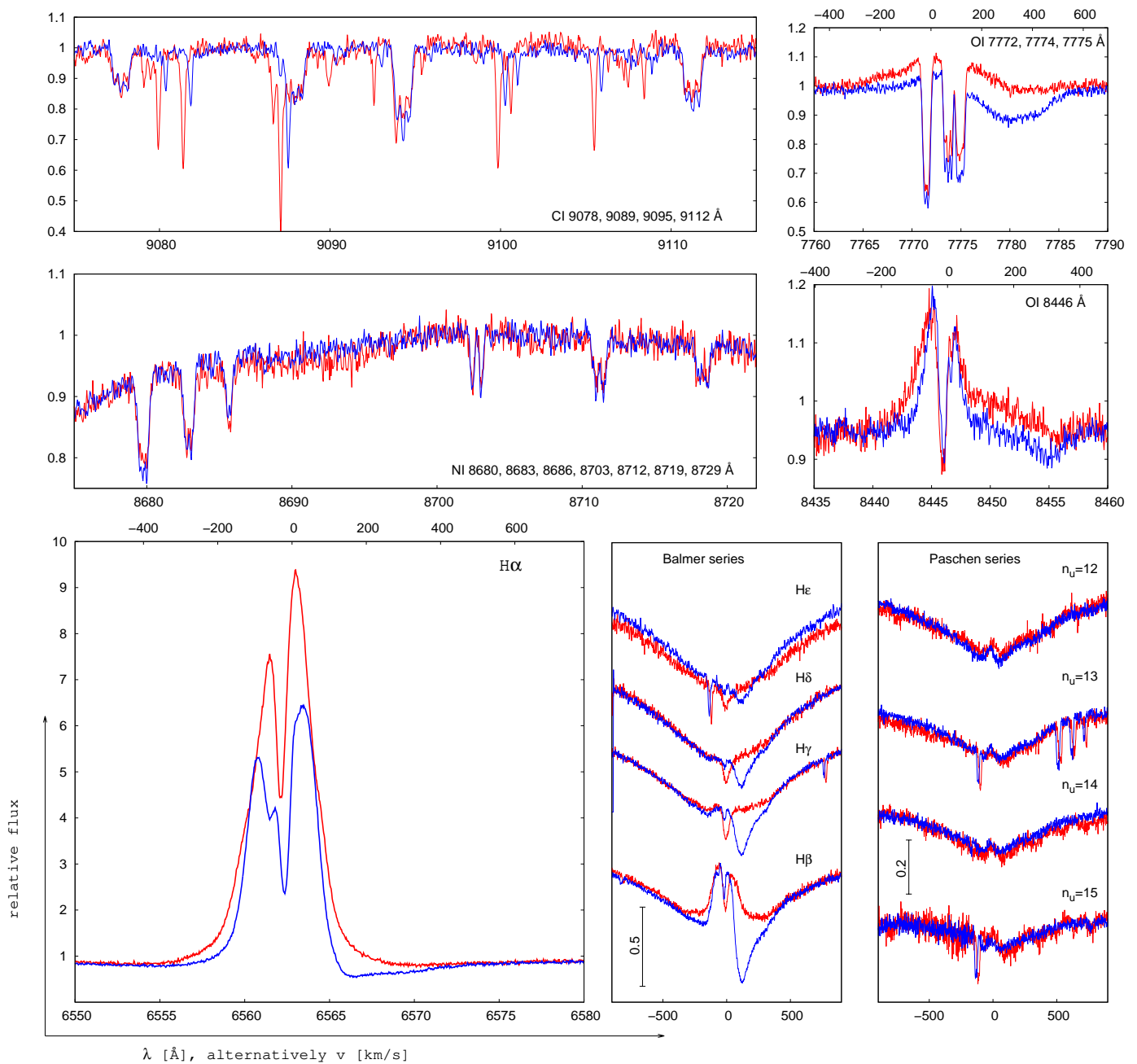}}
  \caption{
   Comparison of the selected parts of the spectra for phases showing the red-shifted absorption (CFHT 2012-02-09, blue lines) 
   and a regular spectrum (CFHT 2006-06-08, red lines).    The spectra are corrected on system velocity determined based 
   on the position of [\ion{O}{i}] lines. The area with \ion{C}{i} multiplet is contaminated by the atmospheric lines.
   }
  \label{variability_component}
\end{figure*}

We found a very atypical behavior in the spectral lines of IRAS~17449+2320. Occasionally, very strong red-shifted absorption
appeared in all the Balmer lines and in the \ion{O}{i} triplet 7~772, 7~774, and 7~775~\AA. The \ion{O}{i} 8~446~\AA \, 
line also shows a slight difference in the red wing. Only minor changes can be seen in the center of the Paschen series. 
The shape of other spectral lines remains unchanged. We show the \ion{C}{i} and \ion{N}{i} multiplet 
in Fig.~\ref{variability_component} to demonstrate that even if the shape of the line profile does not change,  the intensity does change.
This indicates the variability in the stellar continuum. The influence of the continuum intensity is different in 
different parts of the spectrum. 
The spectrum with the red-shifted absorption of the H$\alpha$ and \ion{O}{i} lines shows deeper metal lines at longer wavelengths 
from $\sim 6\,300$~\AA, while shallower at shorter wavelengths from this limit.

\subsection{Variability of Balmer lines}
\label{Balvar}

\begin{figure}[h!]
  \includegraphics[width=\linewidth]{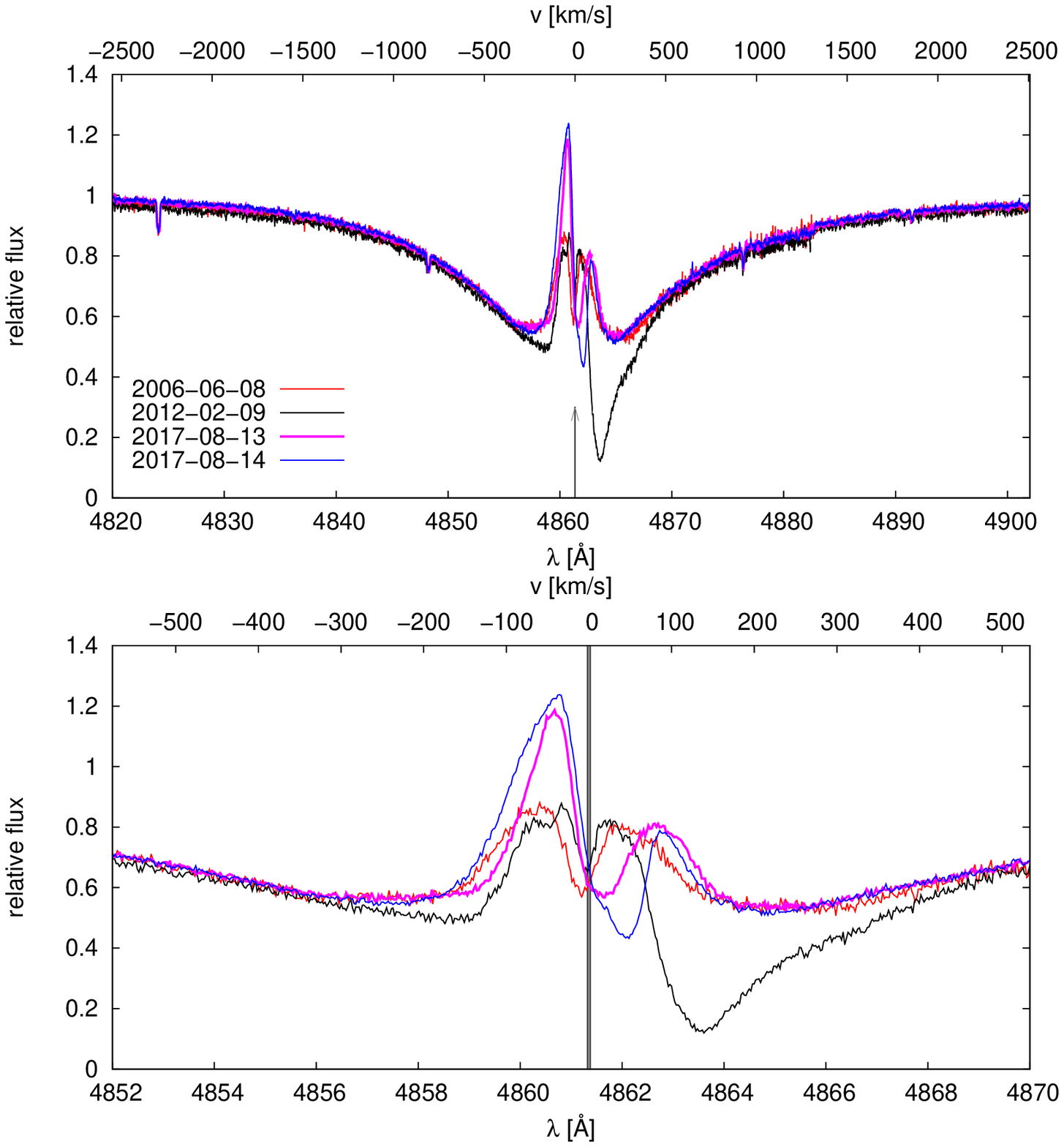} 
 \caption{Selected H$\beta$ line profiles from CFHT. Upper panel shows remarkably stable wings lasting more than decade, while 
     the bottom panel shows the variable core. The grey arrow on the upper panel and grey vertical line on the bottom panel 
     denote the system velocity position. 
   }
 \label{Hbvarfig}
\end{figure}

\begin{figure}[h!]
  \resizebox{\hsize}{!}{\includegraphics{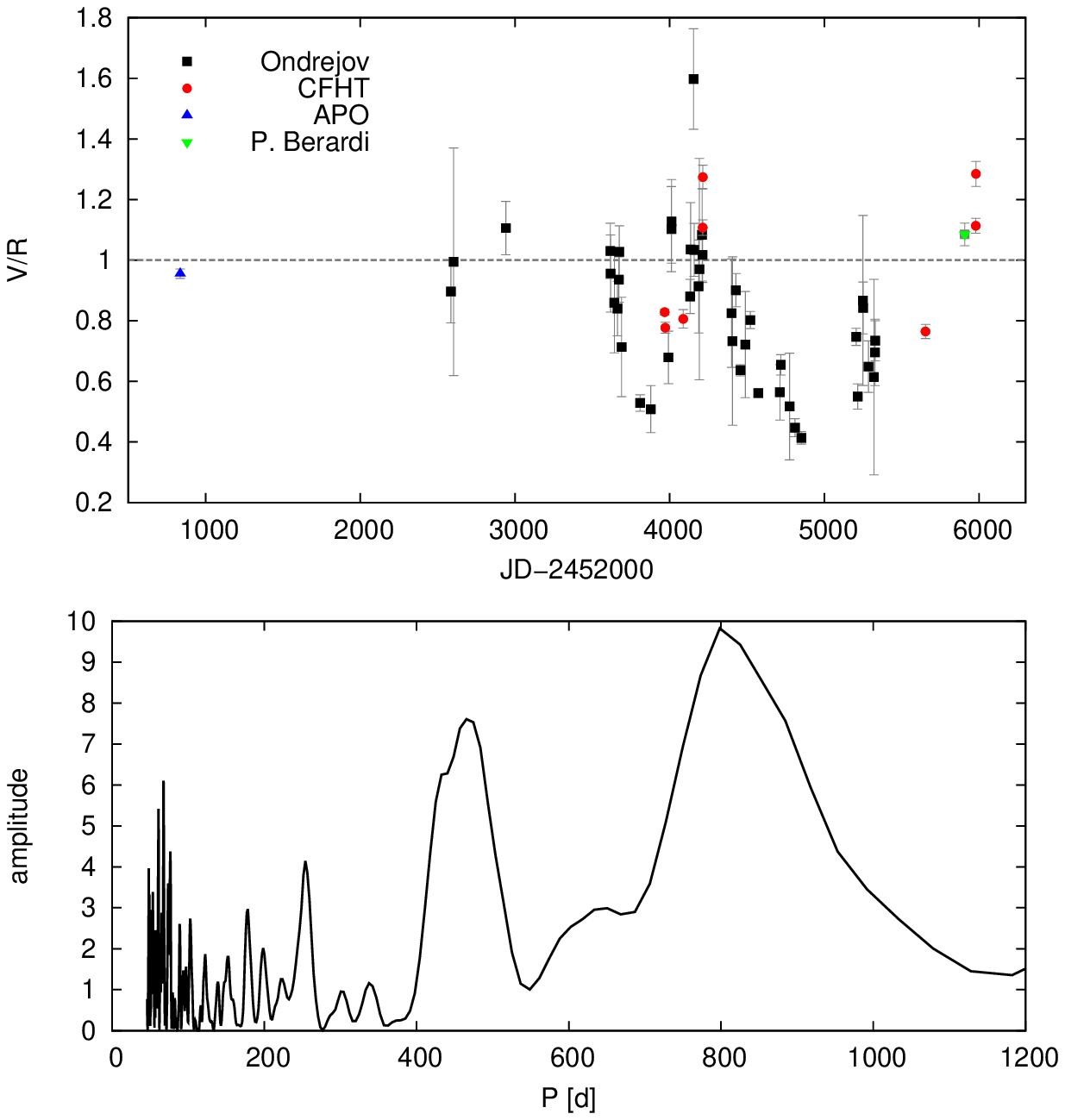}}
  \caption{
   $V/R$ changes of the H$\alpha$ line and its periodogram. The horizontal line in the upper panel shows 
   the $V/R = 1$ level. Most FS~CMa stars have $V/R$ ratios below this line.
   }
  \label{VR_fig}
\end{figure}

Higher members of the Balmer series are variable on time scale of hours; however, only their central emission region
up to  $\sim \pm 500$~\kms) varies. The absorption wings remain stable for years, as we show in
 Figs.~\ref{variability_component} and \ref{Hbvarfig}. The strong variability is shown in the ratio of the intensity of 
the violet and red peaks of the H$\alpha$ line ($V/R$, Fig.~\ref{VR_fig}). The data show smooth, almost periodic behavior, with 
many data points occuring when the violet peak is greater then the red peak. We used a publicly available tool 
\texttt{PGRAM} on NASA's web page\footnote{https://exoplanetarchive.ipac.caltech.edu/cgi-bin/Pgram/nph-pgram}
because it is based on the Lomb-Scargle method \citep{Press} and, therefore, it is equipped to deal with data that are not equally spaced. The 
period is determined to be 798~days (Fig.~\ref{VR_fig}, bottom panel). We do not add the error to the result, 
because this number can not be consider as a~regular period, but, rather, as 
"some scale of the variability." Figure~\ref{VR_fig} shows a~quasiperiodic tendency rather than a~regular one. This behavior is typical for all 
FS~CMa stars \citep{FSCMa_diag}. The $\sim$800~d ``period'' describes the difference between two well-defined minima.

\subsection{Radial velocities of \ion{Si}{ii} lines}
\label{rvSiII}

To study the RVs, we chose the \ion{Si}{ii} line 6\,347~\AA \, as the best tracer. 
This line is present in every one of our data sets, and it is always in
absorption on the spectra of IRAS~17449+2320; moreover, it is narrow and symmetric. 
On the other hand, it shows Zeeman-splitting in the high-resolution spectra. Unfortunately,
we do not have sufficient temporal covering for sufficiently strong magnetic null lines.
We measured the RVs using the script from \cite{Honza12}. The value of RVs was determined by 
automatic line-mirroring using the least square method. The error of the method contains 
not only the signal-to-noise ratio (S/N), but also effects of possible slight asymmetries or water contamination. 
The measurements are subject to human control, which allows us to reject
unsatisfactory spectra. The Lomg-Scargle method, the box-fitting least-squares method, as well as the \cite{Plavchan08} 
method implemented in \texttt{PGRAM} all showed no distinctive peak. Nevertheless, we used the most
significant peaks to construct the phase diagram using \texttt{PHOEBE} 
\citep[eclipsing binary modeling software][]{PhoebeI, PhoebeII, PhoebeIII}. All obtained solutions
were too affected by the measurement errors and they were very likely unrealistic. 

On the other hand, the scatter of \ond \, data is larger than the error of the individual measurements,
suggesting a more complex behavior. Taking into account a very strong magnetic field, 
\ion{Si}{II} is probably concentrated in some spots at the surface  \citep{Silvester14}.  
Unfortunately, our data quality and temporal covering are not sufficient to reveal the full nature of this behavior.


\section{Analysis}


\subsection{Non-LTE effects}
\label{NLTE}

The central star is the most important source for the determination of the temperature of the circumstellar
 matter leading to the presence of neutral and singly ionized metals.
In addition, the source of the strong UV radiation also affects the level population.

The UV radiation excites the resonance lines, causing multiple scattering in the circumstellar region.
In that situation, the coincidence of the wavelength of individual lines of different elements starts to 
play an important role. Moreover, the velocity gradient in the photosphere and the circumstellar region
do not reach high values, features that support this type of interaction. 
One of the most important factors is the connection of the population of the hydrogen
and oxygen levels through the radiation at L$\beta$ and \ion{O}{i} resonance line at 1~026~\AA. 
The downward cascade creates emission in the 8~446~\AA\, line, emission in the forbidden line at 5~577~\AA, and
the forbidden doublet $\lambda\lambda$ $6300, 6364$~\AA.
As a~consequence, the variability of the H$\alpha$ and [\ion{O}{i}] $\lambda\lambda$ $6300, 6364$~\AA\, lines
show the same behavior. A similar situation was found in another FS~CMa star, HD~50138 \citep{Terka17}.
This effect has to be taken into account in the analysis of the temporal behavior of IRAS~17449+2320. These conditions allow the creation 
of autoionization lines. We found weak emission line at 4~355~\AA \,
and weak absorption lines 3~952, 4~918, 5~573, and 7~157~\AA. In addition, UV radiation can also form the emission wings,
which is the case of the strongest \ion{C}{i} lines (Fig.~\ref{CIgrotrian}), two \ion{N}{i} lines
(Fig.~\ref{NIgrotrian}), and \ion{He}{i}  6~678~\AA.

\ion{We find that Fe}{i} and especially \ion{Fe}{ii} lines play an important role.
A huge number of spectral lines coincide very frequently with lines of other elements. Therefore,
the UV pumping of iron lines is very strong. The followup cascades create the emission wings of iron lines. 
The UV absorption of iron group elements is very strong in FS~CMa stars leading to the creation of so called ``iron curtain''
\citep{Ivan-Pariz}. However, this effect is not so strong in IRAS~17449+2320 because there are no iron lines fully 
in emission and only a few of them, which are directly connected with the UV transitions, showing emission wings
(\ion{Fe}{ii} $\lambda\lambda$ 4923.92, 5018.44, 5169.03~\AA).

\subsection{Magnetic field}

\begin{figure*}[h!]
  \resizebox{\hsize}{!}{\includegraphics{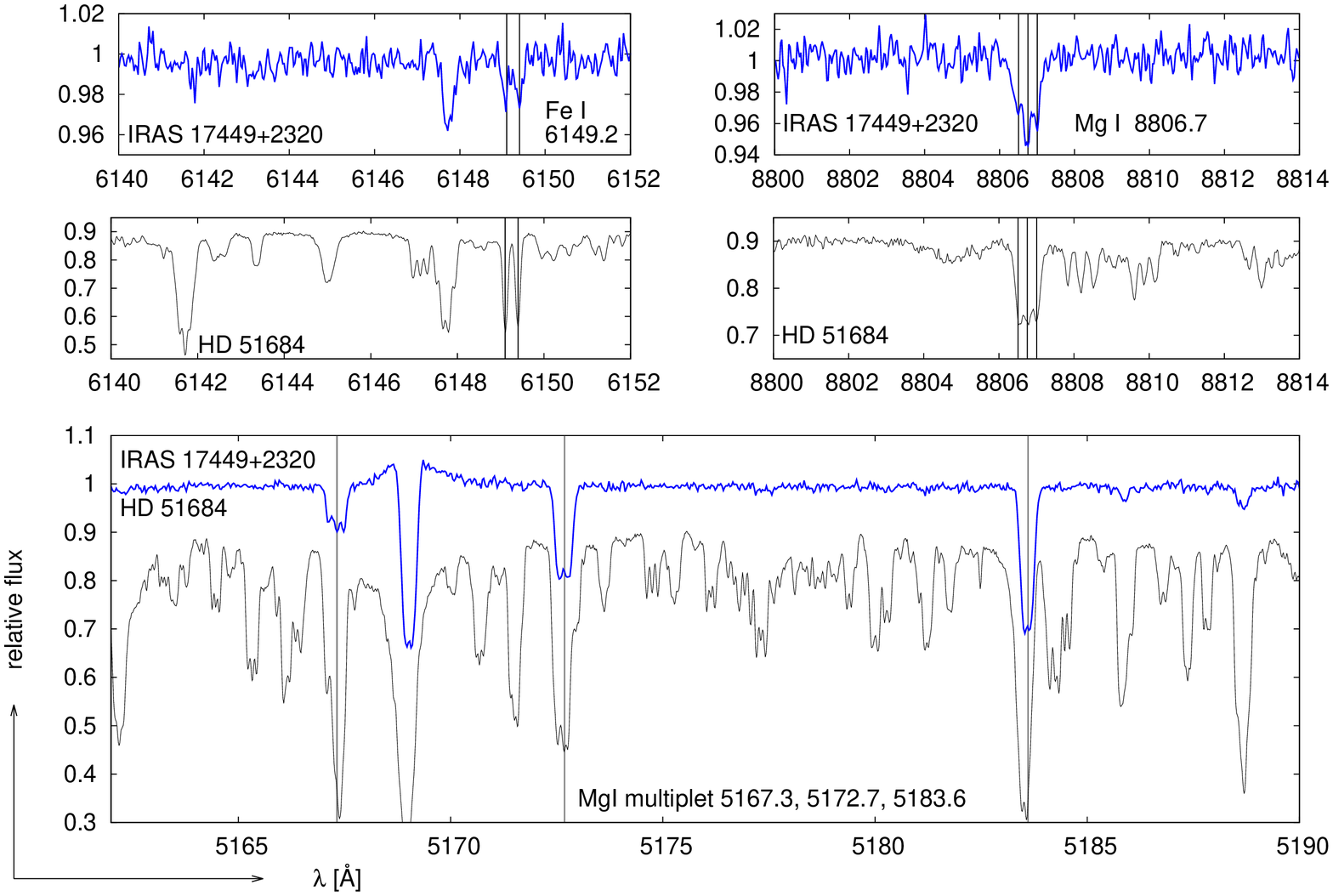}}
  \caption{
    Comparison of selected \ion{Fe}{i} and \ion{Mg}{i} lines of IRAS~17447+2320 and the Ap star HD~51684 (F0p) with the value
    of the mean magnetic field modulus: $6~027\pm50$~G \citep{Mathys17}.
   }
  \label{mag_HD_51684}
\end{figure*}

IRAS~17447+2320 shows a clear Zeeman splitting of metallic lines of \ion{C}{i}, \ion{N}{i}, \ion{O}{i}, 
\ion{Mg}{i}, \ion{Ti}{ii}, \ion{Fe}{i}, and \ion{Fe}{ii,} which makes it possible for application in the case of a straightforward
measurement of the magnetic field. We also found Zeeman splitting in the lines of \ion{Cr}{i} (4~254.35~\AA), 
\ion{Cr}{ii} ($\lambda \lambda$ 4~242.38, 4~252.62, 4~558.650, 4~565.740, 4~836.22, 5~237.32185~\AA), and
\ion{Ce}{ii} ($\lambda \lambda$ 4~770.91, 5~045.12~\AA), however, these lines are weak and noisy in our spectra leading to large errors in the magnetic field determination.
To guess the size of the magnetic field, we compare the spectrum of IRAS~17447+2320 with that of
the Ap star HD~51684 (Fig.~\ref{mag_HD_51684}), which has a mean magnetic field modulus of $6~027\pm50$~G \citep{Mathys17},
and the Am star $o$~Peg (Fig.~\ref{mag_omi_Peg} in the Appendix), with a magnetic field of about 2~kG \citep{Mathys_omi_Peg}. 

We measure the separation of the Zeeman components using a Gaussian profile fitting in \texttt{IRAF} task 
\texttt{splot} on the CFHT spectrum taken on 2017 August 13. The results are summarized in Table~\ref{B_tab}. The 
error is only a formal error of the fit. The corresponding value of the mean magnetic field modulus
$|{\bf B}|$ in G is calculated based on \citep[e.g.,][]{Adelman74}:
\begin{align}
 \Delta \lambda = 4.667 \times 10^{-13} \, g_{\rm{eff}} \, \lambda_{0}^{2} \, |{\bf B}|.
\end{align}
Here, $\Delta \lambda$ in \AA \, is the value of the Zeeman shift, namely, half of the separation of the Zeeman components for doublets,
$g_{\rm{eff}}$ is the effective Land\'{e} factor, and $\lambda_{0}$ is the laboratory wavelength of the line in \AA \,
without the magnetic field presence.

The number of \ion{Fe}{ii} lines with a large split and wide wavelength range is appropriate for a study
of the radial gradient of the magnetic field based on the different line formation regions of 
individual lines. Indeed, the data show a~slightly decreasing value of  $|{\bf B}|$ with increasing 
energy of the lower level (Fig.~\ref{mag_FeII} in the Appendix) with increasing wavelength.
However, this guess has to be examined on a larger sample of high quality data to suppress the measurement errors. 
Currently, a more reasonable value is the arithmetic average of the data, which is $6.2\pm 0.2$~kG.
We show the value of the mean magnetic field modulus for two other epochs for which we have lower S/N 
spectra in the Table~\ref{B_tab_extended}.

\begin{table}
 \caption[]{Mean magnetic field modulus ${\bf |B|}$.}
 \label{B_tab}
 \begin{tabular}{llllc}
   \hline \hline
\multicolumn{1}{c}{$\lambda$} & $g_{\rm{eff}}$& \multicolumn{1}{c}{$\Delta \lambda$}      &  \multicolumn{1}{c}{${\bf |B|}$}                 &  Ref.  \\ 
\multicolumn{1}{c}{(\AA)}     &             & \multicolumn{1}{c}{  (\AA)}               &  \multicolumn{1}{c}{(kG) }             & ($g_{\rm{eff}}$)  \\ \hline
\hline
\multicolumn{5}{l}{\ion{C}{i}} \\ 
9061.4347  &  1.501  &   $ 0.454 \pm 0.006 $  &  $ 7.89 \pm 0.11 $& 1 \\ 
9078.2819  &  1.501  &   $ 0.415 \pm 0.008 $  &  $ 7.19 \pm 0.13 $& 1 \\ 
9088.5097  &  1.501  &   $ 0.405 \pm 0.007 $  &  $ 7.00 \pm 0.11 $& 1 \\ 
9094.8303  &  1.501  &   $ 0.454 \pm 0.005 $  &  $ 7.83 \pm 0.08 $& 1 \\ 
9111.8016  &  1.501  &   $ 0.405 \pm 0.007 $  &  $ 6.96 \pm 0.11 $& 1 \\ 
\hline
\multicolumn{5}{l}{\ion{N}{i}} \\ 
8216.34    &  1.601  &  $ 0.367      \pm 0.010 $  &  $ \,\,\,7.3\,\,\,  \pm  0.2  $ & 2 \\ 
8594.00    &  0.715  &  $ 0.13\,\,\, \pm 0.02\,$  &  $ \,\,\,5.5\,\,\,  \pm  0.8  $ & 2 \\ 
8683.403   &  0.875  &  $ 0.203      \pm 0.012 $  &  $ \,\,\,6.6\,\,\,  \pm  0.4  $ & 2 \\ 
8703.247   &  1.001  &  $ 0.324      \pm 0.008 $  &  $ \,\,\,9.14       \pm  0.22 $ & 2 \\ 
8711.703   &  1.268  &  $ 0.28\,\,\, \pm 0.01\,$  &  $ \,\,\,6.33       \pm  0.22 $ & 2 \\ 
8718.837   &  1.344  &  $ 0.26\,\,\, \pm 0.02\,$  &  $ \,\,\,5.52       \pm  0.4  $ & 2 \\ 
8629.24    &  1.348  &  $ 0.340      \pm 0.014 $  &  $ \,\,\,7.3\,\,\,  \pm  0.3  $ & 2 \\ 
\hline
\multicolumn{5}{l}{\ion{O}{i}} \\ 
7771.94    & 1.084   &  $ 0.216 \pm 0.003$   &  $ 7.05 \pm 0.09$ & 2 \\ 
7774.17    &  1.835  &  $ 0.376 \pm 0.003$   &  $ 7.26 \pm 0.07$ & 2 \\ 
\hline
\multicolumn{5}{l}{\ion{Mg}{i}} \\ 
5172.6843 &  1.877   &  $ 0.116     \pm 0.005$   &  $ 4.93    \pm 0.21$ & 2 \\ 
5183.6042 &  1.376   &  $ 0.089     \pm 0.003$   &  $ 5.1\;\, \pm 0.2$  & 2 \\ 
8806.757  &  1.000   &  $ 0.26\,\,\,\pm 0.02 $   &  $ 7.3\;\, \pm 0.6$  & 2 \\ 
\hline
\multicolumn{5}{l}{\ion{Ti}{ii}} \\ 
4287.89   &  1.50    &  $ 0.07 \pm 0.02$   &   $5.4 \pm 1.4$ & 3 \\ 
\hline
\multicolumn{5}{l}{\ion{Fe}{i}} \\ 
4271.1535  &  1.0     &  $ 0.07 \pm 0.02 $  &   $8.2 \pm 1.9$ & 3 \\
5232.9400  &  1.261   &  $ 0.08 \pm 0.03 $  &   $5.0 \pm 1.7$ & 4 \\ 
\hline
\multicolumn{5}{l}{\ion{Fe}{ii}} \\ 
4122.6591 &  1.005   &  $ 0.065     \pm 0.011  $ & $  \;\,8.2 \;\,\pm  1.4  $ & 5 \\ %
4273.3201 &  1.938   &  $ 0.11\;\,  \pm 0.01 $   & $  \;\,6.7\;\, \pm  0.6  $ & 5 \\ %
4303.17   &  1.221   &  $ 0.069     \pm 0.005  $ & $  \;\,6.49    \pm   0.42$ & 5 \\ %
4576.3330 &  1.200   &  $ 0.069     \pm 0.006  $ & $  \;\,5.9\;\, \pm  0.5  $ & 6 \\ 
4582.8297 &   1.867  &  $ 0.11\;\,  \pm 0.02   $ & $  \;\,6.2\;\, \pm  0.9  $ & 5 \\ %
4620.5128 &  1.333   &  $ 0.09\;\,  \pm 0.02   $ & $  \;\,7.0\;\, \pm  1.4  $ & 6 \\ 
4629.3311 &  1.333   &  $ 0.069     \pm 0.005  $ & $  \;\,5.2\;\, \pm  0.4  $ & 6  \\
4923.9212 &  1.845   &  $ 0.106     \pm 0.003  $ & $  \;\,5.1\;\, \pm  0.12 $ & 5 \\ 
5018.4358 &   1.853  &  $ 0.162     \pm 0.003  $ & $  \;\,7.4\;\, \pm   0.2 $ & 5 \\ %
5169.0282 &  1.077   &  $ 0.096     \pm 0.002  $ & $  \;\,7.2\;\, \pm   0.2 $ & 5 \\ %
6149.2460 &  1.35    &  $ 0.165     \pm 0.020  $ & $  \;\,6.9\;\, \pm  0.9  $ & 7 \\ 
6247.5570 &  1.181   &  $ 0.08\;\,  \pm 0.02   $ & $  \;\,3.6\;\, \pm  0.9  $ & 5 \\ 
\hline
 \end{tabular} 
\tablefoot{The line wavelengths are adopted from the NIST database \citep{NIST_ASD} with 
the exception of \ion{Fe}{ii} lines, for which van Hoof's line list is used \citep{Peters_line_list}.
The effective Land\'{e} factor ($g_{\rm{eff}}$) is presented in the second column, and its data source is given
in the last column. The value of the Zeeman shift together with its formal error are summarized in the
third column, and the corresponding magnetic field modulus in kG is shown in the fourth column. 
}
\tablebib{References to the value of effective Land\'{e} factor $g_{\rm{eff}}$
(1)~\cite{Wolber70};
(2)~\cite{Fischer07};
(3)~\cite{Aslanov};
(4)~\cite{Lozitsky};
(5)~NIST;
(6)~\cite{Mikulasek};
(7)~\cite{Nesvacil04}.
}
\end{table}


\subsection{Stellar parameters}
\label{parameters_Jana}

We used for the spectral fitting the code \texttt{PYTERPOL} 
\footnote{https://github.com/chrysante87/pyterpol}
written by J. Nemravov\'{a} \citep{Jana16}. We used the 
synthetic grids 
\texttt{OSTAR} \citep{Lanz03},
\texttt{BSTAR} \citep{Lanz07},
\texttt{POLLUX} database \citep{POLLUX}, 
\texttt{ATLAS12} \citep{ATLAS12}, and
\texttt{AMBRE} \citep{AMBRE}. 
Even if the code works 
well, the fitting of IRAS~17449+2320 spectra is not straightforward. The first complication arises 
from the presence of the magnetic field. Most spectral lines are split or deformed. The magnetically null lines, 
usually used for the accurate spectral type determination, are too weak or missing in this star. This 
leads us to choose only a few intervals with narrow spectral lines (Fig.~\ref{fit_fig}). It was also necessary 
to avoid the contamination of the continuum by the circumstellar matter 
in the IR region. Therefore, we used only the part of the spectra up to 5~318~\AA.
However, more serious problem is caused by the presence of the hot continuum.
The strong UV radiation (Sect.~\ref{UV_sec}), variability (Sect.~\ref{red_absorption}), 
and the presence of autoionization lines (Sect.~\ref{line_identification_sec})
reveal this additional radiation. We did not detect spectral lines from this source, but 
a~strong continuum changes the line intensities significantly from the short- to 
the long-wavelength region. As the source is only the continuum in our region from near UV to near IR, 
an accurate determination of its temperature and SED in the relative spectra is not possible.  

We obtained the best (Table \ref{pyterpoltab} and Fig.~\ref{fit_fig}) fitting of a~binary with a~hot and cold component 
with the solar metalicity. The temperature of the secondary is about 51\,000~K and its rotation 
velocity reaches the value of 800~\kms. Such a huge value of the rotation velocity means that there are no 
detectable spectral lines of the hot star in the fitted regions. Since we are dealing with a star 
with very strong magnetic field surrounded by the circumstellar material, it would be better to talk
rather about the hot continuum source than about the secondary component. As we detected the signatures 
of the material infall (Sect.~\ref{resonance_sec}), we should take into account also the possibility
that infalling material creates a hot spot on the surface on magnetic poles of the B-type star.

There is no error estimate for the temperature in this table. The reason is that 
51\,000~K is the lower limit of the temperature of the hot source. As we set higher and higher
temperature guesses for the input parameters, the fit was more and more improved, but also the temperature of
the primary star was lower and lower. Therefore, the value of $T_{\rm{eff}}$ ($\sim$ 11~100~K) 
determined for the primary star is only its upper limit. The $\log~g$ 
and $v_{\rm{rot}}$ of the primary star have to be determined well. We also tested the chemical composition
of the primary star using Kurucz's ALTAS grid. The best fit was obtained for the 
solar one. 
                          
\begin{table}[h!]
 \caption[]{Spectral fitting}
 \label{pyterpoltab}
 \begin{tabular}{lll}
\hline \hline
                                   & primary & secondary / hot source \\ \hline
$T_{eff}$ (K)                       & $\leq$ 11040   & $\geq$ 51513  \\
$\log g$ ; $g$  (g$~ \rm{cm}^{-2}$) & 4.1     &  4.0 \\
$v_{\rm{rot}}$ (\kms)                & 9.1     & 800 \\
$lr$                                &  0.57   & 0.43  \\ \hline
 \end{tabular}
\tablefoot{
$lr$ is the relative contribution of bolometric fluxes of the companions to the total bolometric flux. 
The extreme rotation velocity of the secondary component simulates the additional hot continuum. 
}
\end{table}

\begin{figure}[h!]
  \resizebox{\hsize}{!}{\includegraphics{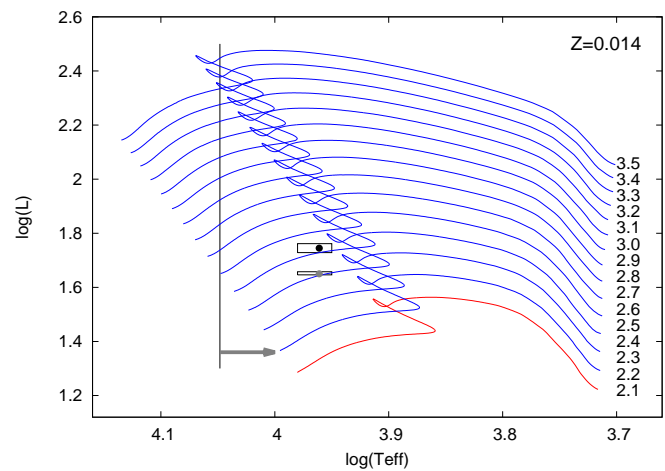}}
  \caption{
    Geneva evolution
     tracks for non-rotating stars without magnetic field for the metalicity $Z=0.014$ \citep{Mowlavi12}. 
    The grey point is the luminosity value derived based on Gaia DR2 without the bolometric correction and dereddening.
    These corrections are applied for the black point. The colour of the evolutionary tracks distinguishes the A- and B-type
    stars. The vertical line shows the upper $T_{\rm{eff}}$ limit determined by the best fit of the spectra. The column on the right-hand side of the
    graph denotes the stellar masses in $M_{\odot}$.
   }
  \label{HRD}
\end{figure}

The most important Gaia EDR3 data \citep{GAIA16_mission,GAIA_EDR3} are summarized in Table~\ref{gaia_data}. 
Based on these data, the HRD is plotted in Fig.~\ref{HRD}. Here, we show the position 
corrected for interstellar reddening \citep{Donnell}. The bolometric correction is applied 
following \cite{Pedersen20}. The values of $R_{v}$ for this star is 3.1, while the $log~g$ value, necessary for the interpolation,
was adopted from the spectral fitting (Table~\ref{pyterpoltab}). The position without the corrections is also plotted 
in Fig.~\ref{HRD} to evaluate the validity of the used approximations. The HRD is plotted with the evolutionary 
tracks of non-rotating and non-magnetic stars. However, this may not be so critical, because \cite{Maeder04} 
showed that magnetic stars rotate as rigid bodies. Therefore, the size of the core, 
main-sequence (MS) lifetimes, tracks, and abundances are closer to the solutions of a non-rotating star
rather than rotating one without the magnetic field.

\begin{table}[h!]
 \caption[]{Gaia EDR3 parameters}
 \label{gaia_data}
\begin{center}
 \begin{tabular}{llll}
\hline \hline
parallax       & \,\, $1.35 \pm 0.02$ mas    & $T_{\rm{eff}}$  &  $9152_{-242}^{+424}$ K$^{\star)}$  \\
distance       & \,\,\, $ 739 \pm 10 $ pc         & $g_{\rm{mean}}$ &  9.953 mag  \\
pma            &  $-2.50 \pm 0.02$ mas/yr &  &\\ 
pmdec          & \,\, $5.91 \pm 0.02$ mas/yr &  & \\ \hline
 \end{tabular}
\end{center}
  \tablefoot{
 $^{\star)}$ The value of $T_{\rm{eff}}$ is taken from the GDR2.
}
\end{table}

\section{Discussion}

\subsection{IRAS 17449+2320's place in FS~CMa group}

Even if the main spectral features of FS~CMa stars are very similar, allowing for the creation of this group,
members do differ from one another in various ways. This is natural because we are dealing with 
objects embedded within very extensive inhomogeneous circumstellar matter, where only a different angle 
of view changes the observed spectral and photometrical properties. Moreover, the presence of a secondary 
component is highly probable. The properties of the secondary as well as the orbital parameters 
affect the observed spectra. In the following, we point to the specific properties of IRAS 17449+2320. 
Since all the members of FS~CMa group are not sufficiently studied, we can mostly compare only with 
the main representatives of FS~CMa group FS~CMa itself, MWC~342, HD~50138, MWC~623, and MWC~728. 

The most remarkable differences are H$\alpha$ $V/R$ changes (\ref{Balvar}). The violet peak is frequently 
larger than the red one and once even 1.6 times. This has not been observed in FS CMa stars. The red peak is 
almost always stronger. Up to now, $V/R>1$ was detected for only one event in HD~50138 
\citep[][$V/R \sim 1.2$]{Terka17} and $V/R \sim 1$ in MWC~728 \citep{MWC728}. 

The H$\alpha$ line of IRAS 17449+2320 shows night-to-night variability. Such a rapid variability of the H$\alpha$ 
line is atypical for FS~CMa stars. We can find a rule in FS~CMa stars -- the absorption lines of 
the B-component are variable on the scale of days and emission on a scale of at least a~week, ruling out 
IRAS 17449+2320. Other lines which show the night-to-night changes are these of \ion{O}{i}, 
and two \ion{Fe}{ii} lines that are connected to UV transitions (Fig.~\ref{variability_night-to-night}). 

On the other hand, fully absorbed \ion{He}{i} lines show no variability. Since \ion{He}{i} lines are formed 
in the deeper regions of the stellar photosphere, we can conclude that IRAS 17449+2320 has a very stable 
photosphere. This is in contrast with the observations of other FS~CMa stars, which show nigh-to-night 
changes, indicating very dynamic inner parts of the photosphere. Here, IRAS 17449+2320  also exhibits a stronger 
difference between the population of the triplet and singlet states. As triplet states have 
lower excitation energy than singlet states, they are highly populated in a low-density medium. 
This stronger overpopulation of \ion{He}{i} lines points to the lower density of the circumstellar matter around IRAS 17449+2320.

\subsection{Spectral type}
\label{spectral_type}

Generally, it is very difficult to exactly determine the stellar parameters of objects with 
extended inhomogeneus atmospheres. Even if FS~CMa stars are slow rotators (i.e., gravity darkening 
could be neglected in these cases), the application of synthetic spectra based on plane-parallel or spherically symmetric models 
has to be taken with caution because of their geometrically-thick and (in most cases) also optically-thick circumstellar disks. 

Based on optical photometry and on the strength of the \ion{He}{i} 4~471~\AA\, and \ion{Mg}{ii} 4~481~\AA\, lines,
\cite{Miroshnichenko_FS_CMa_II} estimated the spectral type A0 for IRAS 17449+2320. 
\cite{Condori19} derived a temperature of $9~200\pm300$~K (A1-A2) and luminosity class II or III from color indexes. 
They also used the equivalent width (EW) ratio of \ion{Mg}{ii} 4~482~\AA\, versus \ion{He}{i} 4~471~\AA\, 
and \ion{He}{i} 4~713~\AA\, versus \ion{Si}{ii} 6~347~\AA\, to determine $T_{\rm{eff}}$  of $9~500 \pm 500$~K (A0 -- A2) and
$10~700 \pm 1000$~K, respectively. Unfortunately, they did not mention the method used for the EW calculation.
Fitting the line profiles by the Voigt function, the error due to the Zeeman splitting would be included. 
Gaia measurements (DR2) give $T_{\rm{eff}}$ $9152_{-242}^{+424}$ K (Table~\ref{gaia_data}).

We show that the exact determination of the stellar parameters is even more difficult for IRAS 17449+2320. 
The best fit by a single star ($T_{\rm{eff}}=12~580$~K, $\log g = 4.2$, and $v_{\rm{rot}}=$11~\kms, solar composition) 
is far from the observed spectrum (see Fig.~\ref{fit_fig}). It is highly probable
that a hot source contributes to the spectrum. Even if we were not able to find the spectral lines of this source,
strong UV radiation (Sect.~\ref{UV_sec}), the autoionization lines (Sect.~\ref{line_identification_sec}),
and especially the spectral variability (Sect.~\ref{red_absorption}) are indicators of the additional hot continuum source. 
We simulate it as a rapidly rotating secondary for the spectral fitting (Sect.~\ref{parameters_Jana}).
As the hot source has no strong spectral lines in the visible, only a lower limit of the $T_{\rm{eff}}$
is possible to obtain. This affects the solution for the ``primary'' star, where it is only possible to determine an
upper limit of $T_{\rm{eff}}=11~040$~K.
Our detailed line identification (Table~\ref{linestab}) rejects the lowest temperatures corresponding 
to the spectral types A1 and A2 because of the lack of many \ion{Fe}{i} lines, which are missing at temperatures 
higher than 10~000~K (e.g. $\lambda\lambda$ 4~107, 4~110, 4~113, 4~114, 4~126, 4~128, 4~135~\AA, etc.). 
Taking into account all the measurements, a~reasonable guess of the spectral type of IRAS 17449+2320
is A0 or B9.

\subsection{Binarity}
\label{binarity}

The only well-defined period ($P = 36.1\pm 0.2$~d) found by \cite{IRAS_Anatoly}
is based on the intensity ratio of the H$\alpha$ emission edges.
They interpreted this period as being due to the orbit of a companion
star. This may be correct since the H$\alpha$ emission edges reflect
the radial velocity changes, rather than the emission strength.
However, generally, the interpretation of the period found based on the emission lines
is not straightforward and the situation is even more complicated in IRAS~17449+2320. 
The complication arises from the magnetic field, which changes the region of the emission formation. 
Instead of being situated in the disk, the emission region might originate 
either from the open magnetic field lines in the polar regions of the star
or from opaque structures, or clouds, that are trapped in the
co-rotating magnetosphere. The presence of such clouds has been recently suggested 
through photometric observations of Landstreet's star by
\cite{Mikulasek20}. If this is the case, then the 36~d period is the stellar rotation period.

\subsection{FS CMa stars as classical Be stars with the magnetic field}

The discovery of the magnetic field in a FS~CMa star opens up a consideration of whether these stars can be 
classical Be stars with a~magnetic field either from the primary or from the secondary component. 
A detailed review of the magnetic field in classical Be stars is given by \cite{Rivi13}. 
Summarizing the published studies, they show that even if there were discoveries of 
a~magnetic field in classical Be stars, it was always close to the detection limit. Eventual 
follow-up studies did not confirm the detections. \cite{Bagnulo12} consider that Be stars do not 
have magnetic fields larger than 100~G. Their data itself cannot exlude the possibility that there 
exists a~Be star with larger magnetic field; however, their analysis shows that 
even if it exists, it would be very exceptional. A very detailed survey of the magnetic field in hot stars 
has been done by the  Magnetism in Massive Stars (MiMeS) consortium. Preliminary results \citep{Wade12} show that even if 
the magnetic field was detected in about 6.5~\% of B-type stars, none of the 58 studied classical Be stars were among them. 
They discussed this finding and demonstrated that it is not an observation or data reduction problem. 

The only Be star with a~magnetic field is $\beta$ Cep. However, \cite{Schnerr06} showed that the magnetic
field is connected with the primary B1 {\sc IV} component and not with the classical Be secondary (B6--8). 
These discoveries reveal that the angular momentum losses during the evolution of a star \citep{Landstreet09} are sufficient
to slow down the rotation speed so that the centrifugal force is not large enough to support the outflow from 
the star to create Be star in a~late MS phase. Thus, $\beta$ Cep points to the possibility that there can be a number of Be stars
with magnetic secondaries. If the separation of the components is sufficiently small to perturb or destroy the Be star disk, 
the observed spectral properties would be similar to that what we observe in FS~CMa stars. However, this is not the case 
for IRAS 17449+2320.

\subsection{Post-MS evolution phase of magnetic Ap stars}
\label{Apmag}

Our discovery of the magnetic field may solve a long-standing problem  \citep{Mathys-Poprad} regarding the nature of post-MS evolution of magnetic Ap stars. 
Some of the lower-temperature FS~CMa stars may be objects that we have been looking for for a long time.

\cite{Landstreet09} presented their results of a study of magnetic Ap stars in open clusters in order to follow the evolution
of their magnetic fields. They found magnetic Ap stars through the entire MS phase, from ZAMS to TAMS. 
However, the strength of the magnetic field strongly decreases with age. This is a~natural consequence of the 
ohmic decay, large-scale hydrodynamic flows, and stellar evolution -- the expansion of the star and reduction of the 
convection core. The ohmic decay time of the fossil magnetic field itself is of the order of $10^{10} - 10^{11}$ years 
\citep{Glagolevskij18}, which is between one and two orders of magnitude longer that the MS lifetime ($10^{8} - 10^{9}$~years). 
As shown by \cite{Landstreet09}, there are still some magnetic stars at TAMS, especially 
among the more massive stars (3-4~M$_{\odot}$), where the surface magnetic field reaches the values of $\sim$500~G. 

The first dredge-up should amplify the rest of the fossil magnetic field and 
cause the amount of matter transported into the circumstellar medium to be larger than in non-magnetic stars. The Ap origin of 
cool FS~CMa stars is partially supported by other observations: 
\textit{i)} FS~CMa stars are located at the end of the MS, or just after TAMS \citep{Miroshnichenko17}; 
\textit{ii)} \cite{Herpin06} discovered magnetic fields in AGB stars:\ the measured values of $B$ were in a range from 0 to 20~G;
\textit{iii)} While classical Be stars rotate at or near the critical velocity,
FS~CMa do not show rapid rotation. The magnetic field reduces the stars' angular momentum. 
The effect is so efficient that the rotation period can reach several decades \citep{Landstreet09}. 

Based on the HRD (Fig.~\ref{HRD}), IRAS~17449+2320 should be a~B-type star. However, 
our spectral fitting gives only the upper limit to the $T_{\rm{eff}}$ because without the UV spectra, 
it is impossible to determine the temperature of the hot source. The Gaia magnitude also has to be affected by this 
hot source; therefore, the luminosity is probably slightly overestimated. 
Previously, \cite{Miroshnichenko_FS_CMa_II} classified this star as A0 and \cite{Condori19} determined the range of
spectral types from A0 to A2 (see Sect.~\ref{spectral_type} for details).
Taking these points into account, the post-main sequence phase of magnetic Ap stars should not be rejected.

\subsection{Mergers}
\label{mergers}

All three previous scenarios suffer from serious problems. The regular periodicity probably reflects the 
stellar rotation rather than the orbital motion. A~classical Be star can not have a~strong magnetic field. 
Rotation will slow down the star during the MS life-time and, thus, it will never become a classical Be star.
Here, IRAS~17449+2320, the primary B-type star, is the one with the strong magnetic field. 
The observed properties point to the post-MS evolution stage of a magnetic Ap star. 
It will fulfill a gap in our observations and models, but the HRD 
(Fig.~\ref{HRD}) is more favorable for the B-type star near the end of its MS life. To explain the observed
properties -- the strong magnetic field, strong IR excess, relatively stable envelope, and the position on the HRD
--  merger theory provides a good fit.

\cite{Schneider20} show that the strong mag. field can be generated during the merger. A strong magnetic field 
can survive the MS evolution because the ohmic decay time is longer than the MS lifetime. The stronger the magnetic 
field, the faster the merged star slows its rotation. The enrichment of the heavy elements in the
photosphere may be detected. However, how large this effect is depends on the age of original stars 
and other factors. For young stars anomalous surface composition does not have to be detected. The merger 
hypothesis naturally explains the large amount of the circumstellar material. Material is ejected: 
\textit{i)} before the merger through the L2 point;
\textit{ii)} during the merger process;
\textit{iii)} after the merger, the resulting star rotates with the critical velocity allowing the creation of the decretion disk; or 
\textit{iv)} in some cases, the Eddington limit may be crossed leading to 
additional mass loss, while some of the ejected material may be re-accreting to the final star.

The merger hypothesis is very promising. The question remains as to whether the merging of binaries has 
sufficiently high probability. \cite{Soker} have presented their assumption that there may be one V838~Mon-like outburst every 
10-50 years in the galaxy. However, the creation of a very bright event is not the only channel 
for the merging process \citep{Soker}. Moreover, they did not take into account the formation of binaries 
\citep[e.g.,][]{Bate02} and evolution of young clusters \citep{Bate19}. Indeed, the simulations of 
recently formed realistic binary-star-rich clusters lead to the formation of "forbidden binaries" that
are characterized by large orbital eccentricities and short periods, leading the companion stars to merge \citep{Kroupa95}.
Recent works on this problem have shown a high ejection rate of massive stars from their birth 
clusters and also reveal that a large fraction of B and O-type stars undergo mergers because of the stellar-dynamical 
encounters in the compact young clusters \citep{Oh15,Oh16,Oh18}. The profusion of mergers, particularly among
massive stars, are likely to be part of the explanation of the elemental peculiarities observed in globular 
cluster stars \citep{Wang20}. In any population, mergers of stars formed in binaries with B-type and also
M-dwarf companion masses are thus likely to be common. This aspect will be addressed in the future through simulations. 
Examples of observed remnants of recent mergers have been monitored \citep{Kaminski15,Tylenda16}.
Moreover, according to current simulations, the binaries ejected from young clusters merge soon after leaving 
the cluster (F. Dinnbier, private communication). This is the natural consequence of the angular momentum change. 

This possible phase of the past evolution has to reveal itself in the proper motion. Indeed, Gaia measurements 
of IRAS~17449+2320 (Table~\ref{gaia_data}) give the space velocities $U=-26.119$, $V=-4.595$, and $W=7.614$~\kms 
\citep{pya}. In comparing these values with the results of an extended study of \cite{Nordstrom04}, we can see that IRAS~17449+2320 has slightly higher
velocities than most of the stars in the solar neighborhood. However, due to the heating of the galactic disk by spiral
arms, or giant molecular clouds,
We also have to take into account the age of stars for the comparison of the space velocities of their stellar sample and IRAS~17449+2320.
According to the HRD (Fig.~\ref{HRD}, bottom panel), IRAS~17449+2320 is about 0.5 Gy old. Among these young stars, 
IRAS~17449+2320 is an outlier. The $W$ velocity, which is the component of the space speed toward the north Galactic pole,
is a particularly large outlier. Based on a more detailed analysis of Gaia DR2 data, \cite{Boubert18} also concluded 
that IRAS~17449+2320 is very likely to be a~runaway star. This brings the strong support for the scenario of the merger binary ejected 
from a young cluster.

\section{Conclusions }

We found the presence of a~magnetic field in IRAS~17449+2320, which is the first detection in FS~CMa type stars. 
The magnetic field is very strong, leading to a clear Zeeman-splitting of many lines, on the basis of which we derived a 
mean magnetic field modulus of $6.2\pm 0.2$~kG. The magnetic field detection changes our view 
of the nature of FS~CMa stars. This opens up the possibility that these objects are classical Be stars, 
the secondary component of which would have a~strong magnetic field and be sufficiently close to the primary 
to disturb or destroy the disk. The magnetic field can hardly be connected with a classical Be star 
because it slows down the rotation speed during the MS evolution. Therefore, the centrifugal force
is too low to support the creation of the decretion disk at the end of the MS. Indeed, no Be stars
with a magnetic field have been detected thus far \citep{Wade12}. Contrary to this assumption, radial velocities and magnetic 
splitting of the lines show that the magnetic field is connected with the primary A/B-type star of IRAS~17449+2320.

Our observations show that IRAS~17449+2320 is a slightly atypical FS~CMa star. Moreover, it is among the
cooler FS~CMa stars. Previously, the spectral type from A0 to A2 had been determined 
(\citealt{Miroshnichenko_FS_CMa_II, Condori19}, see Sect.~\ref{spectral_type}). These are hints that we 
may be dealing with a different type of object. Because the spectrum of IRAS~17449+2320 is very similar 
to magnetic Ap stars (e.g.,~Figs.~\ref{mag_HD_51684} and \ref{mag_omi_Peg}), it allows  for
the possibility that this star could be a magnetic Ap star at an evolutionary stage just after TAMS. 
No such star has  been discovered thus far, although some magnetic Ap stars with strong magnetic fields 
have also been found to be located at TAMS \citep{Landstreet09}. Contrary to this hypothesis, we have the position of 
IRAS~17449+2320 on the HRD (Fig.~\ref{HRD}), leading us to surmise that we are dealing
with B-type star. However, the observed variability of spectral lines (intensity of which is reducing and amplifying simultaneously)
and the spectral fitting reveal the presence of a hot source ($>50~000$~K). It may be either a secondary component, or
a source connected with the magnetic field. Because there are no spectral lines of the hot component 
in the visible part of the spectra, the spectral fitting gives only the upper temperature limit 
of the primary ($\sim$ 11~000~K). The hot source must also be affecting the Gaia measurements, leading to 
an overestimate of the luminosity. 

A strong magnetic field, strong IR excess, emission lines, as well as forbidden emission lines, 
a~relatively stable envelope, and its position on the HRD may be explained as the features of a merger system. 
A magnetic field, even a very strong one, may have been generated during the merger \citep{Schneider20}. 
Due to the magnetic field, the merger product slows down very quickly, as it is observed in FS~CMa stars. 
The material is ejected before, during, and after the merger, leading to the creation of a~massive disk. 
Current calculations show that the escaped binaries from the clusters have to merge soon after
leaving the cluster. IRAS~17449+2320 may be one of these cases, because its component
of the space velocity toward the north Galactic pole is significantly larger ($W=8.47$~\kms) than
is observed in stars in the solar  neighbourhood \cite{Nordstrom04}. \cite{Boubert18} also classified 
IRAS~17449+2320 as a runaway star based on Gaia DR2 data.

The presence of a hot source points to the binary nature of IRAS~17449+2320, where the secondary is a hot dwarf
because it contributes less than 40\% to the total bolometric flux of the system (Sect.~\ref{parameters_Jana}).
The current binary nature of the system is supported by the periodic variations of the H$\alpha$ emission wing edges 
and EWs (36.1~d, \citealt{IRAS_Anatoly}). On the other hand, the variability of the emission parts of lines, 
especially the H$\alpha$ line, does not actually prove orbital motion in a~binary system. 
Especially in this case, the emission may originate in the plasma trapped by the strong magnetic field above 
the surface \citep{Mikulasek20}.

IRAS~17449+2320 does not show all of the typical features of the main representatives of FS~CMa stars \citep{FSCMa_diag}. 
The absorption lines of He and metals, with the exception of the \ion{O}{i} lines, are remarkably stable. 
The night-to-night variability is observed only in H, \ion{O}{i}, and two \ion{Fe}{ii} lines that are directly 
connected with the UV transitions. This behavior is contrary to what is seen in other FS~CMa stars.
The detection of H$\alpha$ $V/R>1$ in IRAS~17449+2320 is also very rare.

Another phenomenon that has been observed for the first time in FS~CMa stars is the opposite behavior with regard to Balmer and 
\ion{O}{i} lines as compared to other lines. We observed at certain epochs the appearance of the strong red wing 
absorption in Balmer lines and the \ion{O}{i} triplet $\lambda\lambda$ 7~772, 7~774, and 7~775~\AA, as well as \ion{O}{i}
line 8~446~\AA \, (Fig.~\ref{variability_component}). No other line profile is affected, which is very unusual. 
The other lines show only an intensity decrease or increase, depending on the position in the spectra. This indicates 
changes in the continuum radiation. Longward from $\sim 6\,300$~\AA, \, the continuum intensity is slightly lower, 
whereas shortward it is slightly higher. It is possible that in these phases, what we are seeing is the secondary component. Alternatively, 
the stellar surface of the magnetic star, which is not blocked by the circumstellar matter distributed along 
the magnetic field lines close to the poles, is visible in these phases. The rotation is fully responsible 
for the spectral variability in this case. We noticed a larger contrast in the population of singlet and 
triplet states of \ion{He}{i} lines, which points to a~lower density material than that of other FS~CMa stars. 
This is supported also by the great number of H levels. We were able to identify the lines from the 28th level.

IRAS~17449+2320 has strong UV radiation (Table~\ref{GALEX_tab}). Even without knowledge of its UV spectra, 
we can deduce its basic character from the visible lines. Since we do not observe emission wings of metal lines,
which are affected by cascade transitions in FS~CMa stars, there may not be an iron curtain at work here. On the other hand, 
the absence of the iron curtain might only be a geometrical effect. If we do not see the disk almost edge-on, 
absorption of the iron group elements would not be so strong, nor would it be the case for the \ion{Li}{i} lines. 
We detected two episodes of material infall, revealed by the narrow discrete absorption components 
in the resonance lines. Usually, material outflow is detected in FS~CMa stars, but the inflow events 
are not exceptional \citep{FSCMa_diag}. 

From the first spectrum of IRAS~17449+2320, it became obvious that we are dealing with an object that can help reveal the
nature of FS~CMa stars. After obtaining high-resolution spectra, we found a Zeeman-splitting of the magnetically 
active lines. This was a~proof that the magnetic field has to be taken into account in the discussion of the nature of FS~CMa stars. 
The strong magnetic field, observed spectral properties, and variability, taken together with the high space velocity toward the galactic pole, 
point to a merger as the most likely scenario for IRAS~17449+2320.

\begin{acknowledgements}

We appreciate the work of the referees, whose valuable comments helped to improve the paper. 
We would like to thank A.~Miroshnichenko, S.~V.~Zharikov, and F.~Dinnbier for the valuable comments and
P.~Zasche, M.~Wolf, P.~Berardi, and P. \v{S}koda for some spectra. The research of DK is supported by~grant 
GA 17-00871S of the Czech Science Foundation. PK acknowledges support from the Grant Agency of the Czech Republic under
grant number 20-21855S. A. Raj acknowledges the Research Associate Fellowship with order no.
03(1428)/18/EMR-II under Council of Scientific and Industrial Research (CSIR).

This work has made use of data from the European Space Agency (ESA) mission {\it Gaia} 
(\url{https://www.cosmos.esa.int/gaia}), processed by the {\it Gaia} Data Processing and Analysis Consortium 
(DPAC, \url{https://www.cosmos.esa.int/web/gaia/dpac/consortium}). Funding for the DPAC
has been provided by national institutions, in particular the institutions
participating in the {\it Gaia} Multilateral Agreement.

\end{acknowledgements}

\bibliographystyle{aa} 
\bibliography{iras17449_2320_Korcakova} 


\begin{appendix}

\section{Additional figures}

\begin{figure}[h!]
 \centering
 \begin{tikzpicture}[
          scale=5.0,
        ]
    \pgfmathsetmacro{\nasy}{1.0}
    \pgfmathsetmacro{\nastext}{1.0}
    \pgfmathsetmacro{\delx}{4}
    \pgfmathsetmacro{\sirkahladiny}{1}
    \pgfmathsetmacro{\ground}{0mm}
    \pgfmathsetmacro{\tris}{7.48277591}
    \pgfmathsetmacro{\trip}{8.85066275}
    \pgfmathsetmacro{\trid}{9.83329434}
    \draw[-,color = black,very thick]  (0mm,\ground mm) -- node[below]{\scalebox{\nastext}{2s$^{2}$2p$^{2}$}}  (\sirkahladiny mm,\ground mm);
    \draw[-,color = black,very thick]  (\delx mm,\tris mm) -- node[above, xshift=0.5 mm]{\scalebox{\nastext}{2s$^{2}$2p3s}}(\delx mm+\sirkahladiny mm,\tris mm);
    \draw[-,color = black,very thick]  (0mm,\trip mm) --  node[above, yshift = 1mm]{\scalebox{\nastext}{2s$^{2}$2p3p}} (0mm+\sirkahladiny mm,\trip mm) ;
    \draw[-,color = black,very thick]  (2*\delx mm,\trid mm) --   node[above]{\scalebox{\nastext}{2s$^{2}$2p3d}}  (2*\delx mm+\sirkahladiny mm,\trid mm);
    \draw[-{Latex[length=2mm,width=1mm,angle'=90]} , color = black, very thick] (0.5*\sirkahladiny mm, \ground mm)  -- (2*\delx mm + 0.5*\sirkahladiny mm,\trid mm);
    \draw[-{Latex[length=2mm,width=1mm,angle'=90]}, color = black, very thick] (2*\delx mm + 0.5*\sirkahladiny mm,\trid mm) -- (0.5*\sirkahladiny mm, \trip mm);
    \draw[-{Latex[length=2mm,width=1mm,angle'=90]}, color = black, very thick] (0.5*\sirkahladiny mm, \trip mm) -- (\delx mm + 0.5*\sirkahladiny mm,\tris mm);
    \draw[-{Latex[length=2mm,width=1mm,angle'=90]}, color = black, very thick]  (0.5*\sirkahladiny mm, \ground mm) -- (\delx mm + 0.5*\sirkahladiny mm +0.5mm,\tris mm) ;
    \node at (0 mm, 12mm) {\Large \ion{C}{i}};
    \node at (6.2 mm, 6mm) {UV};
    \node at (4 mm, 10mm) {IR};
    \node at (1.2 mm, 7.4 mm) {multiplet 270};
    \node at (2. mm, 4mm) {UV};
 \end{tikzpicture}
\label{CIgrotrian}
\caption{Formation of the strongest \ion{C}{i} lines in the visible from the multiplet 270; $\lambda \lambda$
9~061, 9~062, 9~078, 9~089, 9~095, 9~112~\AA. The level 2s$^{2}$2p3s is strongly populated by the UV resonance transition, 
which allows the absorption lines to the level 2s$^{2}$2p3p. However, the upper level from this transition is also 
populated from the level 2s$^{2}$2p3d pumped by UV radiation. This overpopulation of 2s$^{2}$2p3p creates the emission 
wings of these strongest lines.
 }
\end{figure}
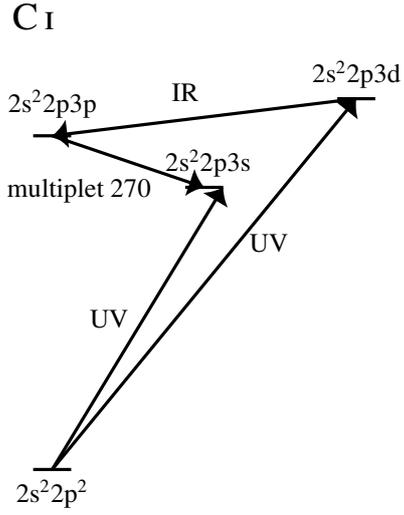

\begin{figure}[h!]
 \centering
 \begin{tikzpicture}[
          scale=5.0,
        ]
    \pgfmathsetmacro{\nasy}{1.0}
    \pgfmathsetmacro{\nastext}{1.0}
    \pgfmathsetmacro{\delx}{4}
    \pgfmathsetmacro{\sirkahladiny}{1}
    \pgfmathsetmacro{\ground}{0mm}
    \pgfmathsetmacro{\dva}{2.3846100}
    \pgfmathsetmacro{\tri}{3.5756182}
    \pgfmathsetmacro{\ctyri}{12.1263786}
    \pgfmathsetmacro{\pet}{10.6899807}
    \pgfmathsetmacro{\sest}{13.0332040}
    \draw[-,color = black,very thick]  (0mm,\ground mm) --  (\sirkahladiny mm,\ground mm);
    \draw[-,color = black,very thick]  (0mm,\dva mm) -- node[left, xshift=-1mm]{\scalebox{\nastext}{2s$^{2}$2p$^{3}\, ^{2}\rm{D}^{0}$}} (\sirkahladiny mm,\dva mm);
    \draw[-,color = black,very thick]  (0mm,\tri mm) --  node[left, xshift=-1mm]{\scalebox{\nastext}{2s$^{2}$2p$^{3} \, ^{2}\rm{P}^{0}$}}(\sirkahladiny mm,\tri mm);
    \draw[-,color = black,very thick]  (0mm,\ctyri mm) -- node[left, xshift=-1mm]{\scalebox{\nastext}{2s$^{2}$2p$^{2}3p \, ^{2}\rm{P}^{0}$}}  (\sirkahladiny mm,\ctyri mm);
    \draw[-,color = black,very thick]  (\delx mm +0mm,\pet mm) -- node[above, yshift=1 mm]{\scalebox{\nastext}{2s$^{2}$2p$^{2}$3s $^{2}$P }}  (\delx mm + \sirkahladiny mm,\pet mm);
    \draw[-,color = black,very thick]  (2*\delx mm +0mm,\sest mm) -- node[right, xshift=1mm]{\scalebox{\nastext}{2s$^{2}$2p$^{2}$3d $^{2}$D}} (2*\delx mm + \sirkahladiny mm,\sest mm);
    \draw[-{Latex[length=2mm,width=1mm,angle'=90]} , color = black, very thick] (0.5*\sirkahladiny mm, \dva mm)  -- (2*\delx mm +0.6*\sirkahladiny mm,\sest mm);
    \draw[-{Latex[length=2mm,width=1mm,angle'=90]} , color = black, very thick] (0.5*\sirkahladiny mm, \tri mm)  -- (2*\delx mm +0.4*\sirkahladiny mm,\sest mm);
    \draw[-{Latex[length=2mm,width=1mm,angle'=90]} , color = black, very thick] (2*\delx mm + 0.5*\sirkahladiny mm,\sest mm)  -- (0.5*\sirkahladiny mm,\ctyri mm);
    \draw[-{Latex[length=2mm,width=1mm,angle'=90]} , color = black, very thick]  (0.5*\sirkahladiny mm,\ctyri mm) -- (\delx mm +0.4*\sirkahladiny mm, \pet mm);
    \draw[-{Latex[length=2mm,width=1mm,angle'=90]} , color = black, very thick] (0.5*\sirkahladiny mm, \dva mm)  -- (\delx mm +0.6*\sirkahladiny mm, \pet mm);
    \draw[-{Latex[length=2mm,width=1mm,angle'=90]} , color = black, very thick] (0.5*\sirkahladiny mm, \tri mm)  -- (\delx mm +0.6*\sirkahladiny mm, \pet mm);
    \node at (1.2 mm, 10.6 mm) {multiplet 75};
    \node at (0 mm, 15mm) {\Large \ion{N}{i}};
    \node at (5 mm, 6mm) {UV};
    \node at (4 mm, 13.2mm) {IR};
    \node at (1 mm, 7mm) {UV};
 \end{tikzpicture}
\label{NIgrotrian}
\caption{
 Formation of the emission wings of lines 8~594.00 and 8~629.24 \AA \, of \ion{N}{i} from the multiplet 75. The UV radiation pumps 
the level 2s$^{2}$2p$^{2}$3d $^{2}$D as well as the level 2s$^{2}$2p$^{2}$3s $^{2}$P, thanks to which the absorption core is observed. 
The level 2s$^{2}$2p$^{2}$3d $^{2}$D is connected straightforward with the ground level by the very strong resonance line. However, 
the energy difference of the levels (corresponding to the $\sim$ 950~\AA) is to high for the UV radiation outgoing from the star. 
The sharp fall around 900 \AA \, is the common property of FS~CMa objects \citep{FSCMa_diag}. Indeed, we found no line corresponding 
to this level in the spectrum.
 }
\end{figure}
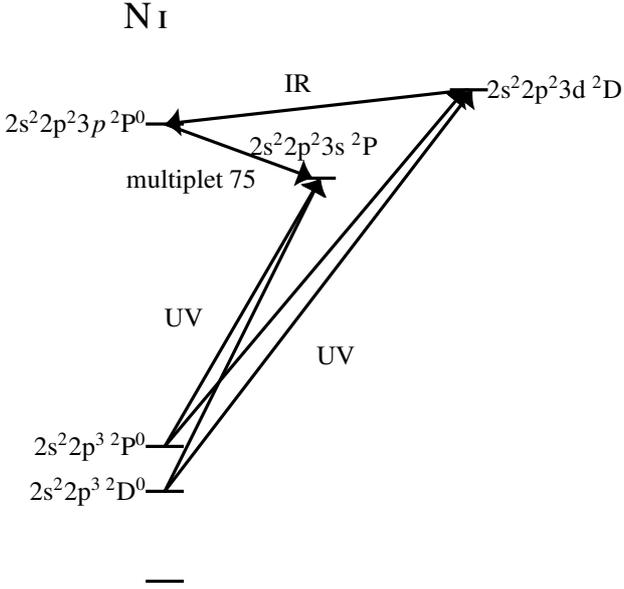

\begin{figure}[h!]
  \resizebox{\hsize}{!}{\includegraphics{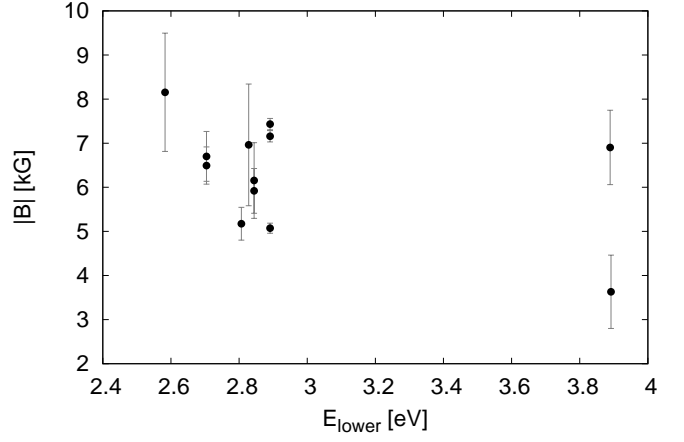}}
  \caption{Mean magnetic field modulus of \ion{Fe}{ii} lines in dependence of the energy of the lower level.
   }
  \label{mag_FeII}
\end{figure}

\newpage
\clearpage
\onecolumn

\begin{figure*}[h!]
  \resizebox{\hsize}{!}{\includegraphics{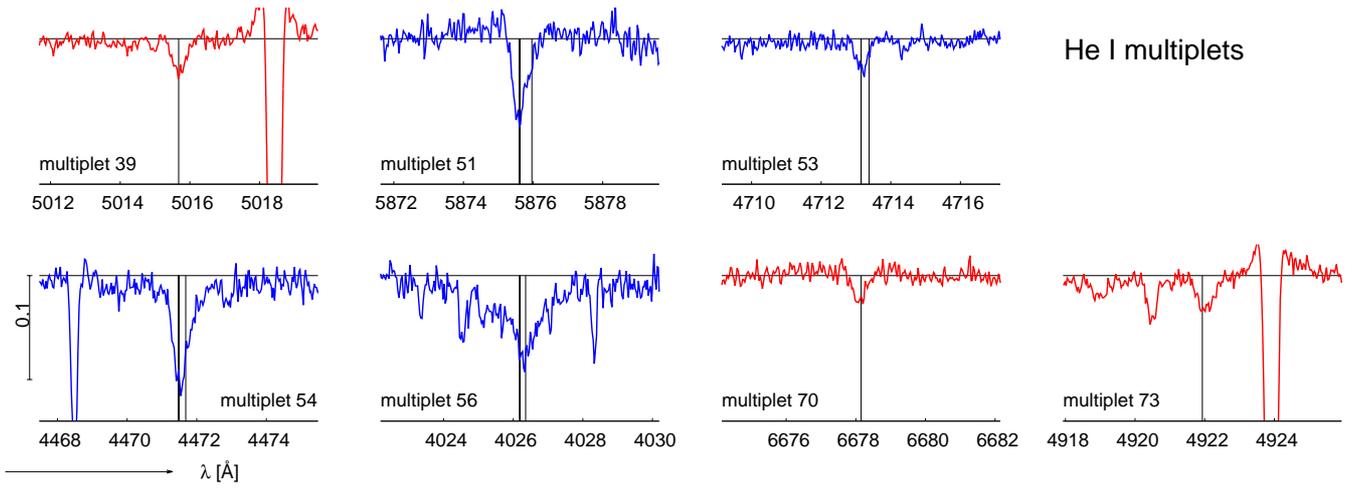}}
  \caption{
   Position, strength, and shape of \ion{He}{i} lines. 
   The vertical lines denote the central wavelength of the lines and the horizontal line the position of
   the continuum. The color distinguishes the singlet (red) and triplets (blue) states.
     All spectral lines are plotted on the same scale, which suppresses the strength of the emission wings of 
     \ion{He}{i} 5\,876 and 6\,678~\AA \, lines. We note that the continuum normalization was carried out very carefully.
   }
  \label{He_identification}
\end{figure*}

\begin{figure*}[h!]
  \resizebox{\hsize}{!}{\includegraphics{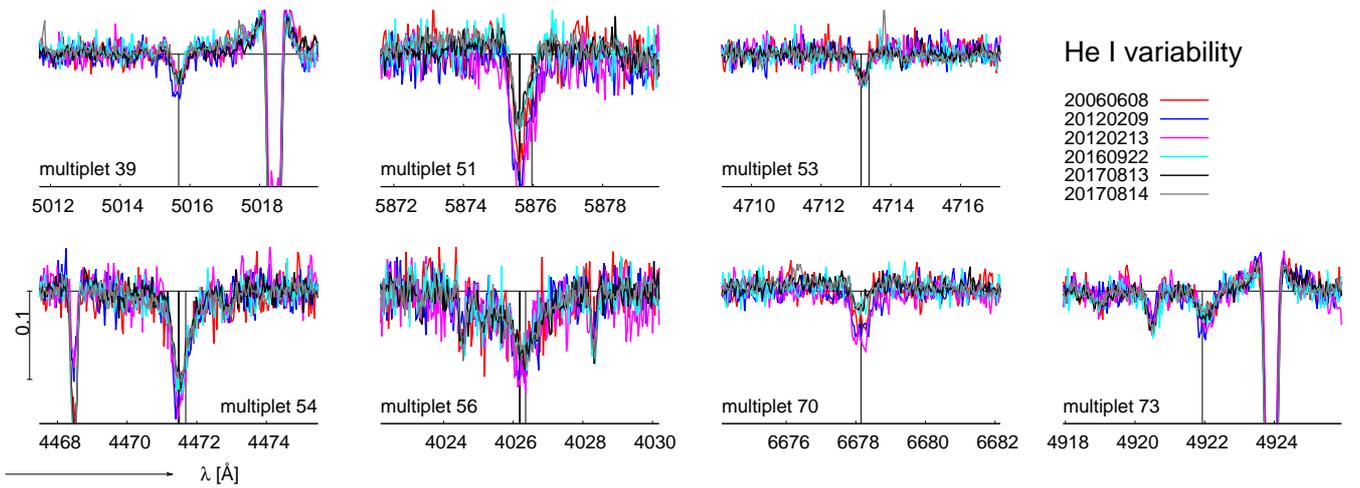}}
  \caption{
Variability of \ion{He}{i} lines. The plotted spectra come from CFHT, which provides the highest available resolution.  
   }
  \label{He_variability_fig}
\end{figure*}

\begin{figure*}[h!]
  \resizebox{\hsize}{!}{\includegraphics{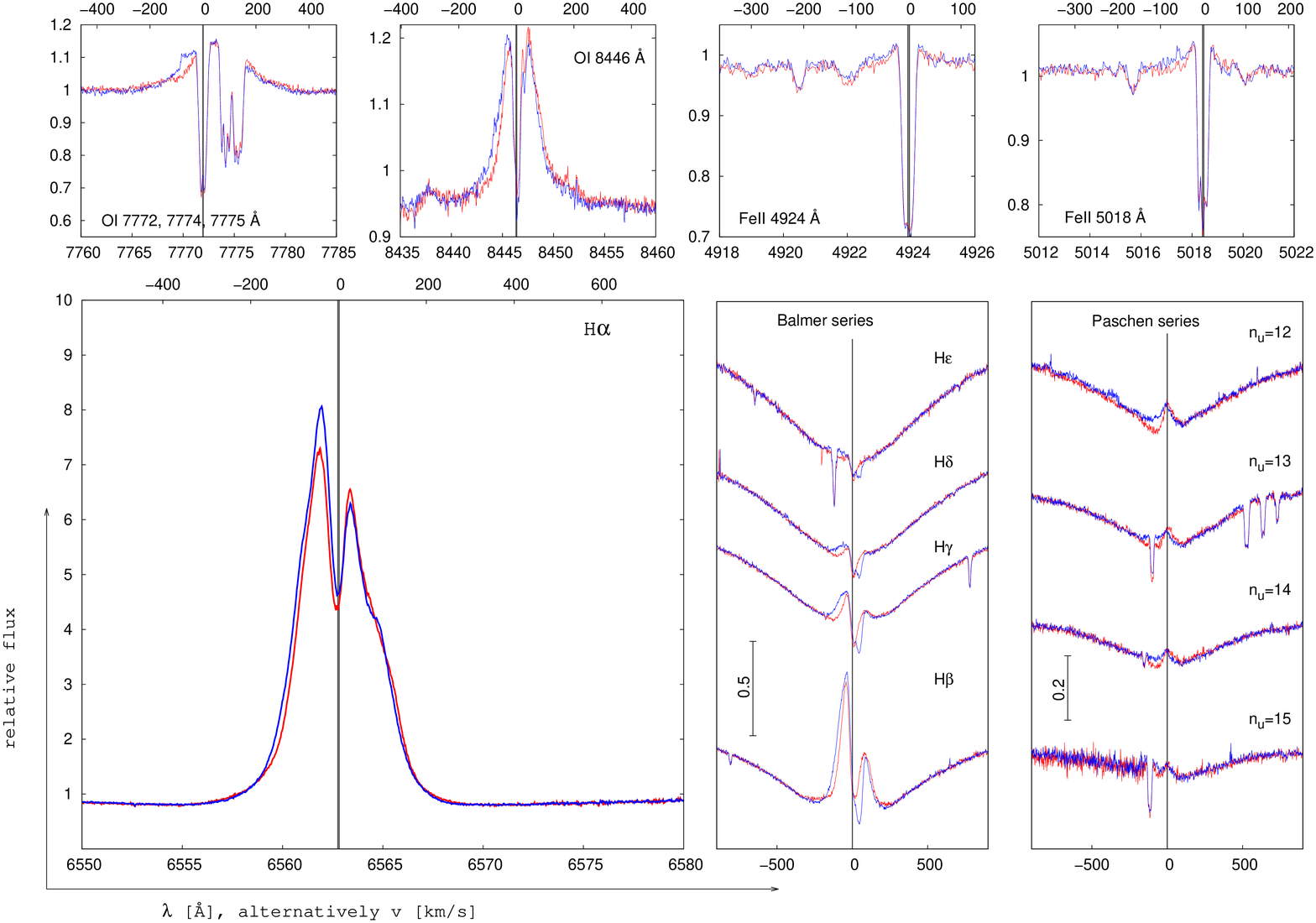}}
  \caption{
  Night-to-night changes demonstrated on the spectra taken at CFHT on 2012-08-13 and 2012-08-14.
  Significant changes are shown only in the hydrogen and oxygen lines. Also shown to be slightly different is the blue 
  wing of the \ion{Fe}{ii} lines 4~924, and 5~018~\AA \, lines. Other lines do not show night-to-night 
  variations. The spectra are corrected on the system radial velocity. The laboratory wavelength
  connected with the system is plotted by the vertical grey line and its thickness corresponds
  to its error.
   }
  \label{variability_night-to-night}
\end{figure*}

\begin{figure*}[h!]
  \resizebox{\hsize}{!}{\includegraphics{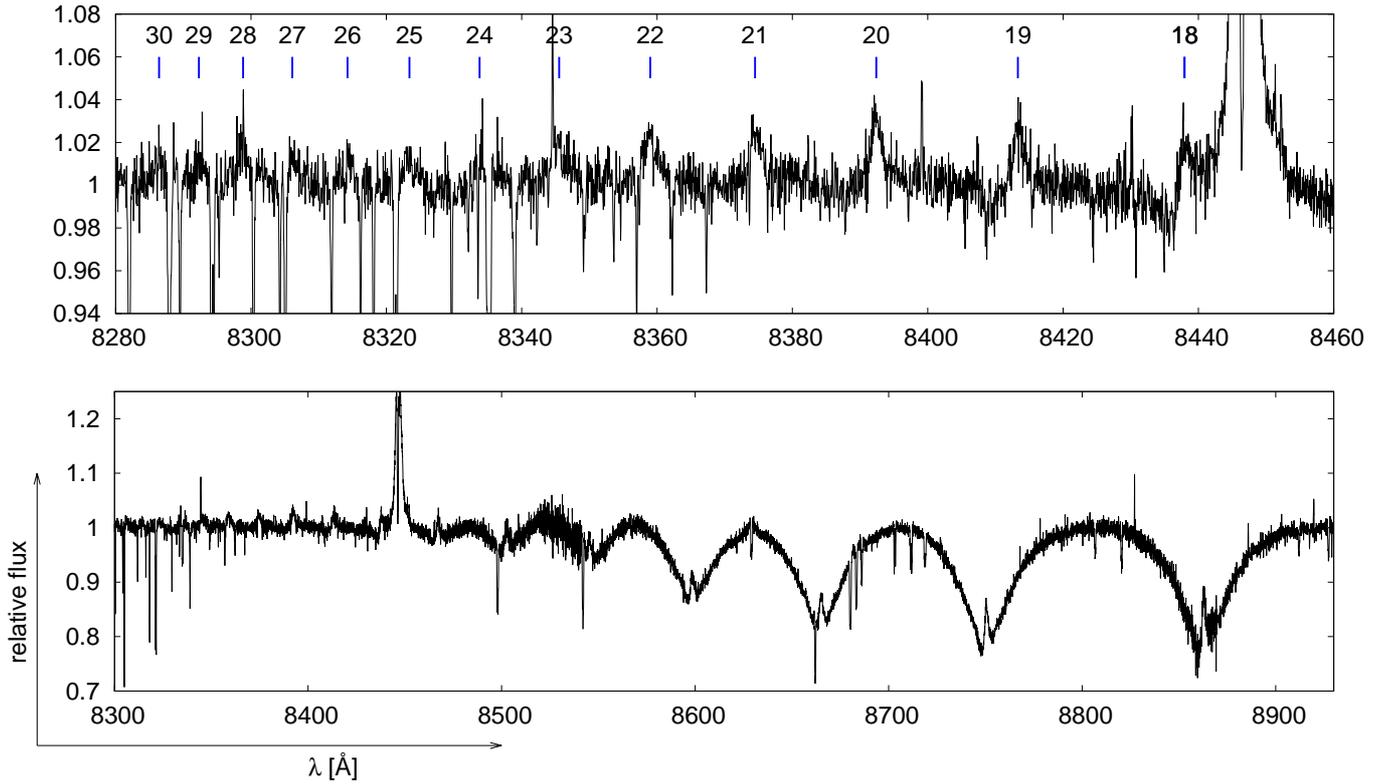}}
  \caption{
    Paschen series. The labels denote the upper level of the transition.  
   }
  \label{paschen}
\end{figure*}

\begin{figure*}
  \resizebox{\hsize}{!}{\includegraphics{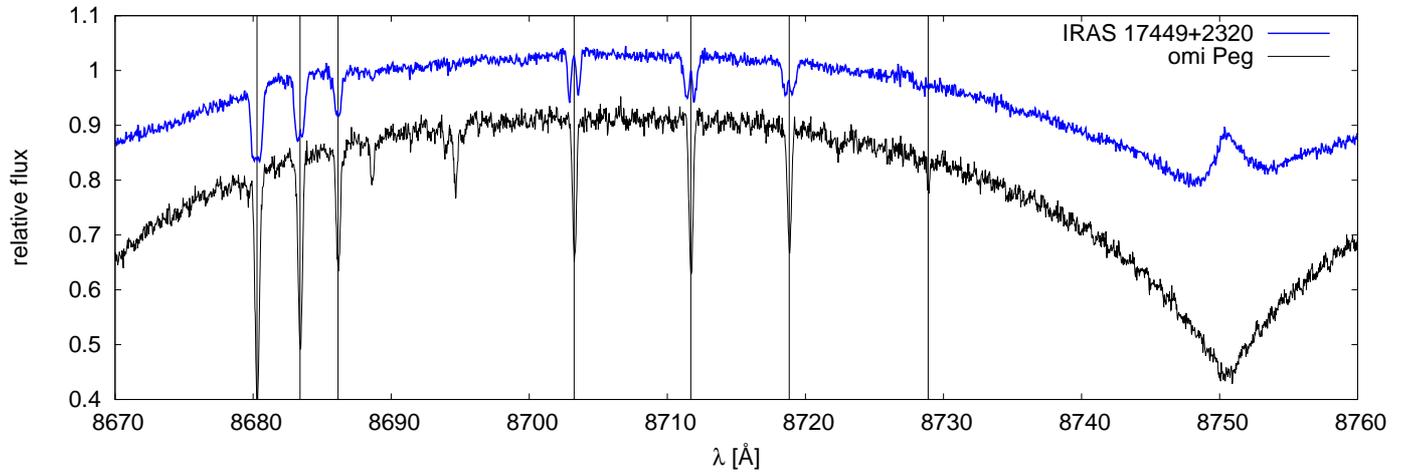}}
  \caption{Comparison of \ion{N}{i} multiplet 60 of IRAS 17449+2320 with a metallic Am star $o$ Peg (A1 IV).
The magnetic field of $o$ Peg reaches the value of 2~kG \citep{Mathys_omi_Peg}. The figure also shows
the Paschen line 8~750.46 \AA.
   }
  \label{mag_omi_Peg}
\end{figure*}

\begin{figure*}
 \label{grotrian_Federico}
  \includegraphics[width=\linewidth]{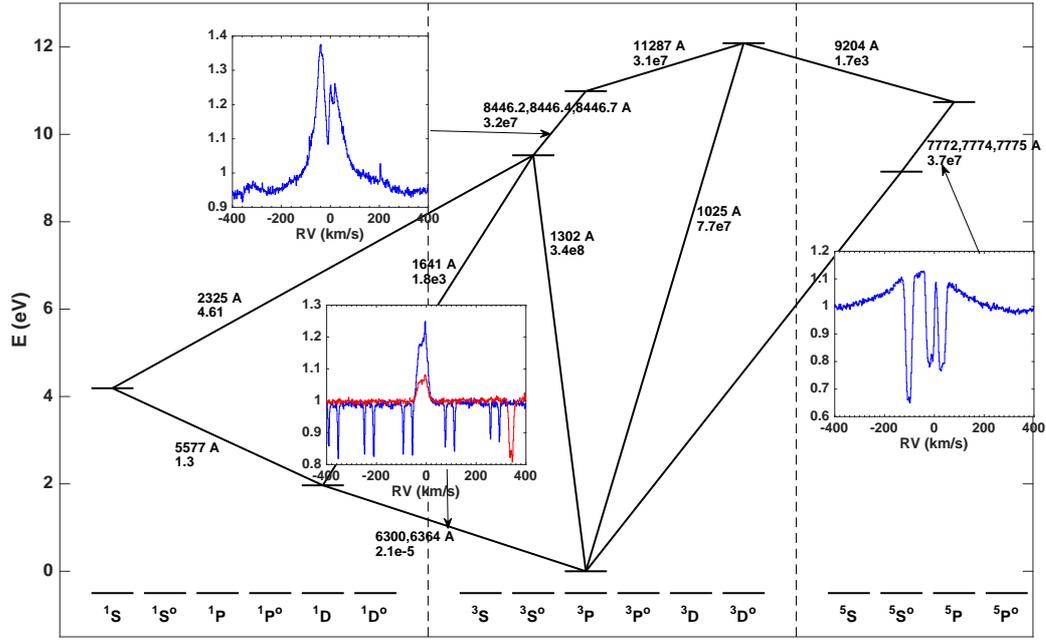} 
  \caption{
    Grotrian of the neutral oxygen.
  }
\end{figure*}

\begin{figure*}[h!]
   \resizebox{\hsize}{!}{\includegraphics{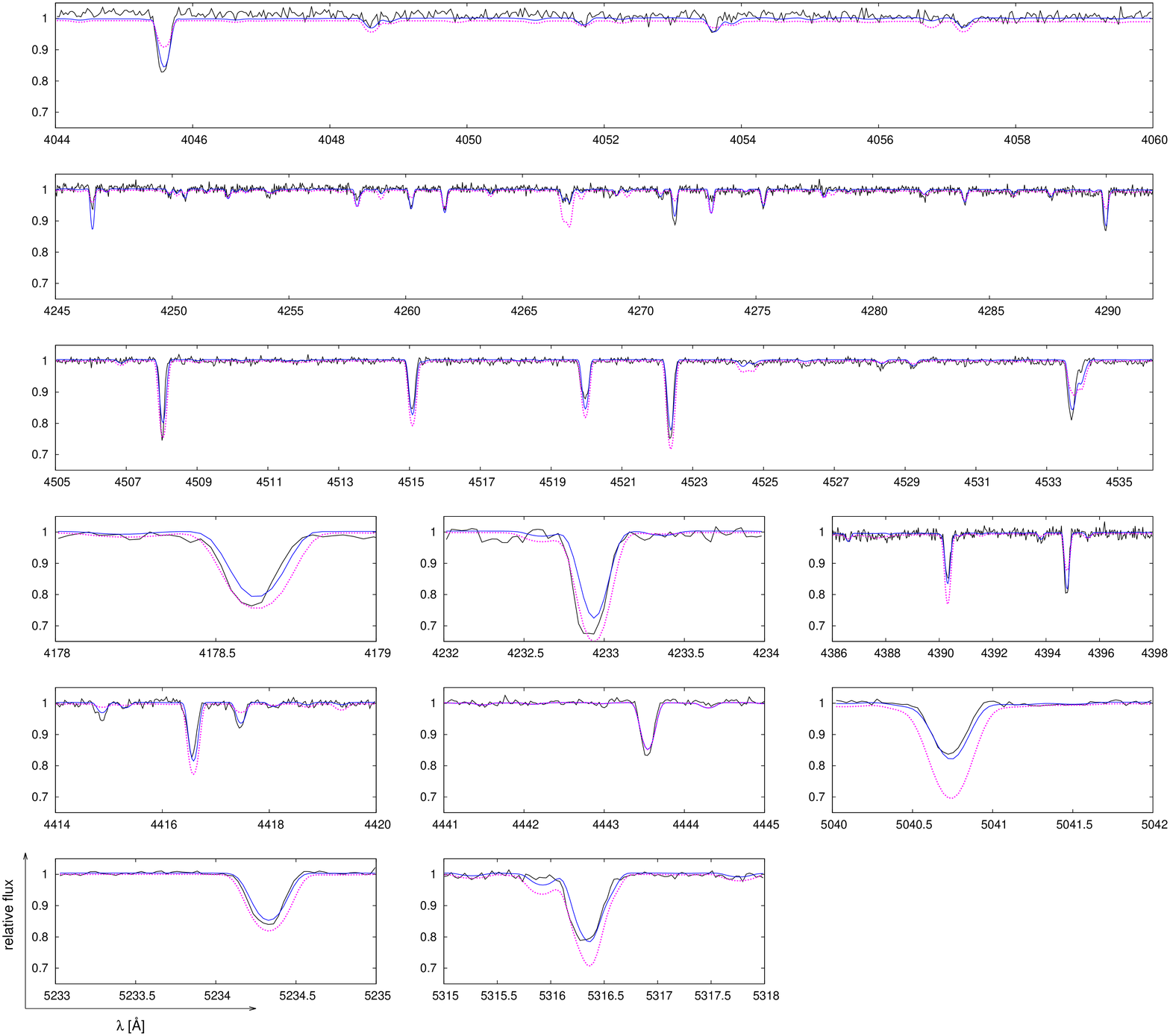}}
  \caption{Spectral fitting. The observed spectrum is plotted by the black line, 
    the blue one is the best fit using \texttt{PYTERPOL} code 
      (see Table~\ref{pyterpoltab}),
      for which the contribution of the hot source was included into the calculations. 
      For the comparison, we also show the best fit by a single star ($T_{\rm{eff}}=12~580$~K, $\log g = 4.2$, and $v_{\rm{rot}}=$11~\kms, solar
      composition) without the contribution of a~hot source, the dotted violet line.
    }
  \label{fit_fig}
\end{figure*}

\begin{figure*}[h!]
   \resizebox{\hsize}{!}{\includegraphics{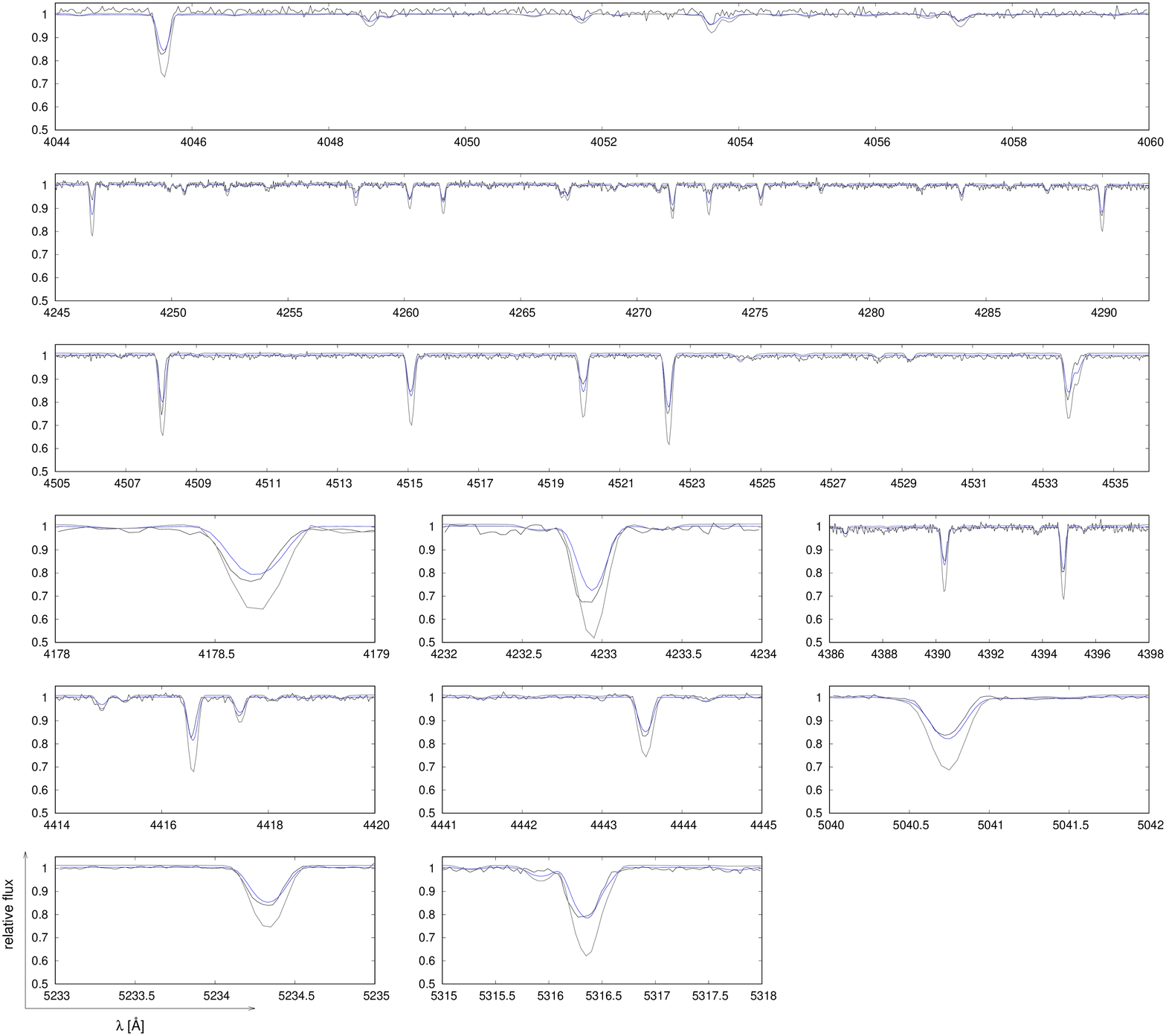}}
  \caption{
    Spectral fitting. The observed spectrum is plotted by the black line, 
    the blue one is the best fit using \texttt{PYTERPOL} code, a star with the contribution of the hot source. 
    The grey line is the individual spectrum of the primary, that is, without the contribution of the hot source. 
   }
  \label{fit_kontrola}
\end{figure*}

\clearpage
\onecolumn
\newpage

\section{Additional tables}

\begin{table}[h!]
 \caption[]{Mean magnetic field modulus ${\bf |B|}$.}
 \label{B_tab_extended}
 \begin{tabular}{lllllc}
   \hline \hline
\multicolumn{1}{c}{$\lambda$} & $g_{\rm{eff}}$& \multicolumn{1}{c}{${\bf |B|}$}    & \multicolumn{1}{c}{${\bf |B|}$}   &  \multicolumn{1}{c}{${\bf |B|}$}    &  Ref.  \\ 
\multicolumn{1}{c}{(\AA)}     &             & \multicolumn{1}{c}{(kG) }          & \multicolumn{1}{c}{(kG) }         &  \multicolumn{1}{c}{(kG) }          & ($g_{\rm{eff}}$)  \\ 
                              &             & \multicolumn{1}{c}{20060608}       & \multicolumn{1}{c}{20120209}       &\multicolumn{1}{c}{20120813}        &         \\ 
                              &             & \multicolumn{1}{c}{$S/N\sim 100$}  & \multicolumn{1}{c}{$S/N \sim 95$ } &\multicolumn{1}{c}{$S/N \sim 200$ } &         \\ \hline                 
\hline
\multicolumn{5}{l}{\ion{C}{i}} \\ 
9061.4347  &  1.501  &                &                & $ 7.89 \pm 0.11 $ & 1 \\ 
9078.2819  &  1.501  & $6.3 \pm 0.2 $ & $ 6.4 \pm 0.2$ & $ 7.19 \pm 0.13 $ & 1 \\ 
9088.5097  &  1.501  &                &                & $ 7.00 \pm 0.11 $ & 1 \\ 
9094.8303  &  1.501  &                & $ 6.3 \pm 0.1$ & $ 7.83 \pm 0.08 $ & 1 \\ 
9111.8016  &  1.501  & $3.2 \pm 0.2$  & $ 6.4 \pm 0.1$ & $ 6.96 \pm 0.11 $ & 1 \\ 
\hline
\multicolumn{5}{l}{\ion{N}{i}} \\ 
8216.34    &  1.601  &  $7.7 \pm 0.1$  & $6.0 \pm 0.2 $ &  $ \,\,\,7.3\,\,\,  \pm  0.2  $ & 2 \\ 
8594.00    &  0.715  &                 &                &  $ \,\,\,5.5\,\,\,  \pm  0.8  $ & 2 \\ 
8629.24    &  1.348  &                 &                &  $ \,\,\,7.3\,\,\,  \pm  0.3  $ & 2 \\ 
8683.403   &  0.875  &  $6.6 \pm 0.4$  & $6.3 \pm 0.3 $ &  $ \,\,\,6.6\,\,\,  \pm  0.4  $ & 2 \\ 
8703.247   &  1.001  &  $9.0 \pm 0.2$  & $8.5 \pm 0.2 $ &  $ \,\,\,9.14       \pm  0.22 $ & 2 \\ 
8711.703   &  1.268  &  $6.1 \pm 0.4$  & $5.7 \pm 0.2 $ &  $ \,\,\,6.33       \pm  0.22 $ & 2 \\ 
8718.837   &  1.344  &                 &                &  $ \,\,\,5.52       \pm  0.4  $ & 2 \\ 
\hline
\multicolumn{5}{l}{\ion{O}{i}} \\ 
7771.94    & 1.084   &  $ 7.0 \pm 0.1$ & $ 6.87 \pm 0.09 $  &  $ 7.05 \pm 0.09$  & 2 \\ 
7774.17    &  1.835  &  $ 7.1 \pm 0.1$ & $ 6.79 \pm 0.06 $  &  $ 7.26 \pm 0.07$  & 2 \\ 
\hline
\multicolumn{5}{l}{\ion{Mg}{i}} \\ 
5172.6843 &  1.877   &  $4.9 \pm 0.21 $ & $4.6 \pm 0.2$  &  $ 4.93    \pm 0.21$  & 2 \\ 
5183.6042 &  1.376   &  $5.0 \pm 0.2 $  & $4.8 \pm 0.2 $ &  $ 5.1\;\, \pm 0.2$   & 2 \\ 
8806.757  &  1.000   &                  &                &  $ 7.3\;\, \pm 0.6$   & 2 \\ 
\hline
\multicolumn{5}{l}{\ion{Ti}{ii}} \\ 
4300.0421 &  1.21 &  $4.8 \pm 0.4 $       &&& 3 \\ 
4301.9225 &  0.83 &  $4.7 \pm 0.8 $       &&& 3 \\ 
4287.89   &  1.50 &                  & &   $5.4 \pm 1.4$ & 3 \\ 
\hline
\multicolumn{5}{l}{\ion{Fe}{i}} \\ 
4271.1535  &  1.0     &    & &  $8.2 \pm 1.9$ &  3 \\
5232.9400  &  1.261   &    & & $5.0 \pm 1.7$  &  4 \\ 
\hline
\multicolumn{5}{l}{\ion{Fe}{ii}} \\ 
4122.6591 &  1.005   &                  &                & $  \;\,8.2 \;\,\pm  1.4  $ & 5 \\ %
4273.3201 &  1.938   &                  &                & $  \;\,6.7\;\, \pm  0.6  $ & 5 \\ %
4303.17   &  1.221   &                  &                & $  \;\,6.49    \pm  0.42 $ & 5 \\ %
4576.3330 &  1.200   &                  &                & $  \;\,5.9\;\, \pm  0.5  $ & 6 \\ 
4582.8297 &   1.867  &                  &                & $  \;\,6.2\;\, \pm  0.9  $ & 5 \\ %
4620.5128 &  1.333   &  $ 6.3 \pm 0.7 $ &                & $  \;\,7.0\;\, \pm  1.4  $ & 6 \\ 
4629.3311 &  1.333   &                  &                & $  \;\,5.2\;\, \pm  0.4  $ & 6  \\
4923.9212 &  1.845   &  $ 4.9 \pm 0.2 $ & $4.6 \pm 0.2$  & $  \;\,5.1\;\, \pm  0.12 $ & 5 \\ 
5018.4358 &   1.853  &  $ 7.4 \pm 0.2 $ & $6.64\pm 0.12$ & $  \;\,7.4\;\, \pm   0.2 $ & 5 \\ %
5169.0282 &  1.077   &                  & $6.5 \pm 0.2$  & $  \;\,7.2\;\, \pm   0.2 $ & 5 \\ %
6149.2460 &  1.35    &                  &                & $  \;\,6.9\;\, \pm  0.9  $ & 7 \\ 
6247.5570 &  1.181   &  $ 5.5 \pm 0.7 $ &                & $  \;\,3.6\;\, \pm  0.9  $ & 5 \\ \hline 
average   &          &  $ 6.0 \pm 0.4$  & $ 5.8 \pm 0.5$ & $  \;\,{\bf 6.2}\;\, \pm  {\bf 0.2}  $                     & \\
\hline
 \end{tabular} 
\tablefoot{The wavelength value is adopted from NIST database \citep{NIST_ASD} with 
the exception of \ion{Fe}{ii} lines for which van Hoof's line list is used \citep{Peters_line_list}.
The effective Land\'{e} factor ($g_{\rm{eff}}$) is printed in the second column and its data source
in the last column. The value of the Zeeman shift together with its formal error is summarized in the
third column and the corresponding magnetic field modulus in kG in the fourth column. 
}
\tablebib{References to the value of effective Land\'{e} factor $g_{\rm{eff}}$
(1)~\cite{Wolber70};
(2)~\cite{Fischer07};
(3)~\cite{Aslanov};
(4)~\cite{Lozitsky};
(5)~NIST;
(6)~\cite{Mikulasek};
(7)~\cite{Nesvacil04}.
}
\end{table}

\clearpage
\newpage

\longtab[2]{
\begin{longtable}{r|r|c|r|r|l} 
\caption{\label{sptab} List of observations. }\\ 
\hline\hline
Date            & HJD & Wavelength range     & Exp. time          & $S/N$ & Observer/project ID \\  
year-month-day  &     & (\AA)                & \hspace*{5mm} (s) &       & \\ \hline 
\endfirsthead
\caption{continued.}\\
\hline\hline
Date            & HJD & Wavelength range     & Exp. time          & $S/N$ & Observer/project ID \\  
year-month-day  &     &                (\AA) & \hspace*{5mm} (s) &       & \\ \hline 
\endhead
\hline
\endfoot
\hline
\endlastfoot
\multicolumn{6}{l}{Canada France Hawaii Telescope; $R=65\,000$ }\\ \hline
2006-06-09  &  2456087.86  & 3820 - 9980  & 6480 & 100  & Manset / 06ad04 \\ 
2012-02-09  &  2455967.13  & 3760 - 9830  & 1800 & 105 & Manset / 12AC12 \\ 
2012-02-13  &  2455971.10  & 3820 - 9890  & 1800 &  95 & Manset / 12AC12 \\  
2016-09-22  &  2457653.71  & 3760 - 9960  & 2134 & 100 & Manset / 16BD94 \\ 
2017-08-13  &  2457978.73  & 3690 - 10300 & 4800 & 170 & Manset / 17BD96 \\ 
2017-08-14  &  2457979.84  & 3700 - 10400 & 3200 & 190 & Manset / 17BD96  \\ \hline 
\multicolumn{6}{l}{Apache Point Observatory; $R=31\,500$ }\\
\end{longtable}
}

\clearpage

\longtab[3]{                                                                                                                                                                                            
\begin{longtable}{l|l|l||l|l|l}                                                                                                                                                                         
\caption{\label{linestab} Line list. }\\                                                                                                                                                                
\hline\hline                                                                                                                                                                                            
Wave length & Element/       & Notes & Wave length & Element/       & Notes \\                                                                                                                          
  (\AA)     &  ionisation    &       & (\AA)     &  ionisation    &       \\ \hline                                                                                                                     
\endfirsthead                                                                                                                                                                                           
\caption{continued.}\\                                                                                                                                                                                  
\hline\hline                                                                                                                                                                                            
Wave length & Element/     &   Notes & Wave length & Element/       & Notes \\                                                                                                                          
  (\AA)     & ionisation   &         & (\AA)       &  ionisation    &       \\ \hline                                                                                                                   
\endhead                                                                                                                                                                                                
\hline                                                                                                                                                                                                  
\endfoot                                                                                                                                                                                                
\hline                                                                                                                                                                                                  
\endlastfoot                                                                                                                                                                                            
3691.551     & \ion{H}{i} H16          &                                         & 4130.872     & \ion{Si}{ii} & A $^{\star}$                                      \\
3697.157     & \ion{H}{i} H15          &                                         & 4130.89      & \ion{Si}{ii} & A     	                                           \\
3703.859     & \ion{H}{i} H14          &                                         & 4132.0579    & \ion{Fe}{i}  & A                                                 \\
3711.978     & \ion{H}{i} H13          &                                         & 4143.8678    & \ion{Fe}{i}  & A                                                 \\
3721.946     & \ion{H}{i} H12          &                                         & 4163.6437    & \ion{Ti}{ii} & A                                                 \\
3734.369     & \ion{H}{i} H11          &                                         & 4167.2712    & \ion{Mg}{i}  & A 	                                               \\
3750.151     & \ion{H}{i} H10          &                                         & 4171.9038    & \ion{Ti}{ii} & A                                                 \\
3759.2915    & \ion{Ti}{ii} & A                                                  & 4173.4512    & \ion{Fe}{ii} & A $^{\star\star}$                                 \\
\end{longtable}                                                                                                                                                                                         
\tablefoot{                                                                                                                                                                                             
\begin{center}                                                                                                                                                                                          
\begin{tabular}{ll|ll|ll} \hline                                                                                                                                                                        
E  &  emission line        & Z & Zeeman split            & iA &  interstellar absorption \\                                                                                                             
A  &  absoprtion line      &  $\quad$ 2Z & double peak   & DIB & diffuse interstellar bands \\                                                                                                          
EC &  emission core        &  $\quad$ 3Z & triple peak   & $^{\star}$  & Ritz wavelength (NIST database) \\                                                                                             
a  &  autoionisation line  &  $\quad$ 4Z & four peaks    & $^{\star \star}$ & Van Hoof's line list \citep{Peters_line_list} \\ \hline                                                                   
\end{tabular}                                                                                                                                                                                           
\end{center}                                                                                                                                                                                            
}                                                                                                                                                                                                       
}

\afterpage{\clearpage}

\end{appendix}

\end{document}